\documentclass[a4paper,11pt]{article}
\usepackage{jheppub} 
\usepackage[utf8]{inputenc}
\usepackage{graphicx}
\usepackage{amsmath}
\allowdisplaybreaks[1] 
\usepackage{hyperref}
\usepackage{cancel}
\usepackage{multirow}
\usepackage{slashed}
\usepackage{amssymb}
\usepackage{booktabs} 
\usepackage{axodraw2}
\usepackage{orcidlink}

\title{\boldmath The Pion Sum Rule beyond the Chiral Limit}

\author[a]{J.~Bijnens\,\orcidlink{0000-0002-1618-2844},\hskip-0.2em}
\author[b]{\,N.~Hermansson-Truedsson\,\orcidlink{0000-0003-3619-2916},\hskip-0.2em}
\author[c]{and A.~Rodr\'{i}guez-S\'{a}nchez\,\orcidlink{0000-0001-7291-2146}}
\affiliation[a]{Division of Particle and Nuclear Physics, Department of Physics, Lund University,
Box 118, \\SE 221 00 Lund, Sweden}
\affiliation[b]{Higgs Centre for Theoretical Physics, School of Physics and Astronomy, The University of Edinburgh, James Clerk Maxwell Building,
Peter Guthrie Tait Road,
Edinburgh,
EH9 3FD}
\affiliation[c]{Departamento de Física. Universidad de Castilla La Mancha. Avenida de Carlos III, s/n, 45004 Toledo}

\emailAdd{johan.bijnens@fysik.lu.se}
\emailAdd{nils.hermansson-truedsson@ed.ac.uk}
\emailAdd{antonio.rsanchez@uclm.es}

\abstract{
The difference between the charged and neutral pion masses
can be predicted from a well-known dispersion relation involving an infinite-energy integral over experimental data, the pion sum rule. This relation, however, holds only in the chiral limit. Here we combine several nonperturbative techniques to determine the form of the physical finite-energy integral, thereby generalizing this sum rule to include effects linear in the light-quark masses. We test the dispersion relation using hadronic tau-decay data.} 

\begin{document}
\maketitle
\flushbottom

\section{Introduction}

At low energies QCD becomes intrinsically nonperturbative: the quark–gluon degrees of freedom that organize the high-energy expansion no longer provide a useful starting point. Yet the same regime admits powerful nonperturbative tools, which allow one to make quantitative statements about hadron dynamics and, in particular, about the interplay between QCD and electromagnetism. Examples of such tools are dispersion theory~\cite{Das:1967it,Moussallam:1997xx,Ananthanarayan:2000ht}, effective field theory~\cite{Gasser:1983yg,Gasser:1984gg} and lattice QCD~\cite{FlavourLatticeAveragingGroupFLAG:2024oxs}. For precision tests of the Standard Model it is often useful to combine them, such as for the recent and successful theoretical prediction of the muon anomalous magnetic moment~\cite{Aoyama:2020ynm,Aliberti:2025beg}.

A well-known nonperturbative dispersion relation, originally derived in Ref.~\cite{Das:1967it}, relates the electromagnetic contribution to the mass difference between the charged and neutral pion, the two lightest hadrons, to an integral over experimental data from the two-pion hadronic threshold up to infinity. In practice, its predictive power is limited by two facts: experimental input is only available up to finite energies, and the relation itself is derived in the chiral limit where quarks are massless. In this work we are interested in elucidating what the corresponding finite-energy integral built from physical data corresponds to within effective field theory, specifically chiral perturbation theory~\cite{Gasser:1983yg,Gasser:1984gg,Urech:1994hd}, and in analyzing the phenomenological consequences.

To this end, we start in section~\ref{sec:pionsrchinf} with a re-derivation of the chiral-limit dispersive relation given in Ref.~\cite{Moussallam:1997xx}, based on chiral perturbation theory for mesons in the presence of electromagnetism~\cite{Urech:1994hd}. The purpose of the re-derivation is to make the paper self-contained and to introduce our notation. In section~\ref{sec:pionsrchfin} we then generalize it to a finite-energy sum rule, going beyond known results. Extending the analysis beyond also the chiral limit is not straightforward, and is the main focus of the paper presented in section~\ref{sec:pionsrphys}. This requires a consistent regularization of divergences associated with QED photon loops at several intermediate steps. It is well-known that some of the electromagnetic low-energy constants appearing in chiral perturbation theory depend on both the gauge parameter and the short-distance matching scale~\cite{Bijnens:1993ae,Bijnens:1996kk,Moussallam:1997xx,Agadjanov:2013lra,Bijnens:2014lea}, which in the latter case is an example of separation-scheme dependence in QCD+QED highly relevant for precision calculations~\cite{Aoyama:2020ynm,FlavourLatticeAveragingGroupFLAG:2024oxs,Aliberti:2025beg}. We carefully track the dependence on these quantities in the matching between chiral perturbation theory and a dispersive integral depending on experimental data, leading to our final result Eq.~(\ref{eq:pionsumrulephys}). Finally, in section~\ref{sec:pionsrphys} we numerically study our matching equation together with hadronic tau experimental data from ALEPH~\cite{Davier:2013sfa}, and compare to the chiral limit sum rule.

The results in this paper pave the way for future phenomenological studies where chiral effects are needed. One example is for isospin-breaking corrections to hadronic observables, which require the inclusion of both electromagnetism and light-quark mass differences. Another avenue is improved determinations of electromagnetic low-energy constants of chiral perturbation theory, which generally are poorly known~\cite{Bijnens:2014lea}.

\section{Reproduction of the pion sum rule in the chiral limit}
\label{sec:pionsrchinf}
A modern re-derivation of the pion sum rule within effective field theory was presented in Ref.~\cite{Moussallam:1997xx}. The starting point is the QCD Lagrangian with three active quark flavors. Following Ref.~\cite{Ecker:1988te}, we promote the charge matrix to left- and right-handed external sources, $Q_L$ and $Q_R$ (set equal at the end),
\begin{equation}\label{eq:lagone}
\mathcal{L}\supset  i \, \bar{q}_L \, i Q_{L}^a\frac{\lambda^a}{2}\slashed{A} q_L + (L\to R) \, .
\end{equation}
Here $A_\mu$ is the photon field and $q_{L/R} = (u_{L/R}, d_{L/R}, s_{L/R})^T$. 
The corresponding low-energy effective field theory, chiral perturbation theory (ChPT) in the presence of these sources, allows for an extra term at $\mathcal{O}(e^2\, p^0)$,
\begin{equation}\label{eq:lageffzero}
\mathcal{L}_{\mathrm{eff}}\supset C\langle Q_R U Q_L U^{\dagger} \rangle \, .
\end{equation}
Here $U$ contains the meson octet fields, $\pi,K,\eta$. Expanding $U$ and setting the external sources to their physical values, $Q_L=Q_R=e\,\mathrm{diag}(2,-1,-1)/3$, one obtains, at this chiral order,
\begin{equation}
\Delta m_{\pi^+}^{2}=  \Delta m_{K^+}^2=\frac{2e^2C}{F^2} \, ,
\end{equation}
where $\Delta m_{P^+}^{2}=m_{P^+}^{2}- m_{P^0}^{2}$, $P=\pi,K$. The $\mathcal{O}(e^2 p^2)$ generalization, $\mathcal{L}_{\mathrm{eff}}=\sum_{i}K_i\, \mathcal{O}_{i}$, was obtained in Ref.~\cite{Urech:1994hd} and introduces additional low-energy constants $K_i$. These are further discussed in section~\ref{sec:eftLR}.

One simple way to link both Lagrangians is to require that the derivatives of the generating functional, $W$, agree. We thus define, decomposing $Q_{L(R)}=\sum_a Q_{L(R)}^a\frac{\lambda^a}{2}$,
\begin{equation}
\Pi_{LR}^{Q}(q)\equiv 4 \int d^{4}x \, e^{iqx} \, \frac{\delta^2 W}{i\delta Q_L^{3}(x)\, i\delta Q_R^3(0)}\Bigg{|}_{Q_{L(R)}\to 0} \, ,
\end{equation}
and evaluate this object in both theories. In the full theory this gives
\begin{equation}
\Pi_{LR}^{Q}(q)= -4i \int d^{4}x \, e^{iqx} \, \langle 0 | T\{\bar{q}_L(x) \,  \frac{\lambda^3}{2}\slashed{A}(x) q_L(x)  \, \bar{q}_R(0) \,  \frac{\lambda^3}{2}\slashed{A}(0) q_R(0)\} |0\rangle \, .
\end{equation}

We may now contract the photon fields. Working in an arbitrary $R_\xi$ gauge, associated with the gauge-fixing term
\begin{equation}
\mathcal{L}_A =-\frac{1}{2\xi}(\partial^\mu A_\mu)^2 \, ,
\end{equation}
the photon propagator is given by\footnote{In Ref.~\cite{Moussallam:1997xx} infrared (IR) divergences were regulated by a photon mass. Here we use dimensional regularization also for IR divergences.}
\begin{equation}
-iD_{\mu\nu}(p)=\frac{-i}{p^2}\left[g_{\mu\nu} + (\xi -1)\frac{p_\mu p_\nu}{p^2} \right] \, .
\end{equation}
One finds
\begin{equation}\begin{aligned}
\Pi_{LR}^{Q}(q)&=4i\int d^{4}x \, e^{iqx}\, [-iD^{\mu\nu}(x)] \langle 0 | T\{J_{L,\mu}^{3}(x) J_{R,\nu}^{3}(0)\}| 0 \rangle   \\
&= 4 i \, \int \frac{d^{4}k}{(2\pi)^4} [-i D^{\mu\nu}(k)] \int d^{4}x \, e^{i x(q-k)} \langle 0 | T\{J_{L,\mu}^{3}(x) J_{R,\nu}^{3}(0) \}| 0 \rangle  \, \\
&= 4 \int \frac{d^4k}{(2\pi)^4}   [-i D^{\mu\nu}(k)] \, \Pi^{3}_{LR,\mu\nu}(k-q)   \\
&=  \int \frac{d^4k}{(2\pi)^4}   [-i D^{\mu\nu}(k)]\,  \Pi^{3}_{V-A,\mu\nu}(k-q)   \, .
\end{aligned}\end{equation}
Here
\begin{equation}
\Pi^{3}_{LR,\mu\nu}(p)\equiv i\int d^{4}x\,e^{ipx}\,\langle 0|T\{J^{3}_{L,\mu}(x)J^{3}_{R,\nu}(0)\}|0\rangle \, ,
\end{equation}
and parity implies $\langle VA\rangle=\langle AV\rangle=0$, so that $\Pi^{3}_{V-A,\mu\nu}\equiv \Pi^{3}_{VV,\mu\nu}-\Pi^{3}_{AA,\mu\nu}=4\,\Pi^{3}_{LR,\mu\nu}$. Here $J_{L,\mu}^3\equiv\bar{q}_L\frac{\lambda^3}{2}\gamma_{\mu} q_L$ and similarly for $L\to R$. In the isospin limit $\Pi^{3,\mu\nu}_{V-A}=\frac{1}{2}\Pi_{ud,V-A}^{\mu\nu}$ as defined in Ref.~\cite{Braaten:1991qm}, and whose imaginary part can be obtained from hadronic tau decay data. Next we Lorentz-decompose the correlator
\begin{equation}\begin{aligned}
\Pi^{\mu\nu}(k)&=(-g^{\mu\nu}k^2+k^{\mu}k^{\nu})\Pi^{(1)}(k^2) +k^{\mu}k^{\nu}\Pi^{(0)}(k^2) 
\\
&=(-g^{\mu\nu}k^2+k^{\mu}k^{\nu})\Pi^{(1+0)}(k^2) +g^{\mu\nu}k^2\Pi^{(0)}(k^2) \, ,
\end{aligned}
\end{equation}
 Here and in the following we use the simplified notation $\Pi\equiv\Pi_3^{V-A}$. One has at zero external momentum $q$
\begin{equation}
\Pi_{LR}^{Q}(0)=-i\int \frac{d^{4}k}{(2\pi)^4}\left[-3\Pi^{(1+0)}(k^2)+(3+\xi)\Pi^{(0)}(k^2)\right] \, .
\end{equation}

The next step in our derivation is to perform a Wick rotation. Assuming the usual analyticity properties, we take $k^0\to i k_E^{0}$ so that $d^4k=i\,d^{4}k_E$ and $k^2\to -k_E^2$. In this way
\begin{equation}
\Pi_{LR}^{Q}(0)=\int \frac{d^{4}k_E}{(2\pi)^4}\left[-3\Pi^{(1+0)}(-k_E^2)+(3+\xi)\Pi^{(0)}(-k_E^2)\right]\, .
\end{equation}

Next we re-express the integral in spherical variables, using that the integrand does not depend on the angular part. Defining $Q^2\equiv k_E^2$, one has
\begin{equation}\label{eq:LRUVd4}
\Pi_{LR}^{Q}(0)=\frac{1}{(4\pi)^{2}}\int_0^{\infty} dQ^2 \, Q^2\left[-3\Pi^{(1+0)}(-Q^2)+(3+\xi)\Pi^{(0)}(-Q^2)\right] \, .
\end{equation}
Now, in the chiral limit, conservation of the octet axial current implies that $Q^2\Pi^{(0)}(-Q^2)=0$ and $\lim_{Q^2\to \infty} Q^4 \Pi^{(1+0)}(-Q^2)=0$, which guarantees that $\Pi^{(1+0)}(-Q^2)$ satisfies an unsubtracted dispersion relation,
\begin{equation}
\Pi^{(1+0)}(-Q^2)=\frac{1}{\pi}\int^{\infty}_0 \frac{ds}{s+Q^2}\mathrm{Im}\Pi^{(1+0)}(s) \, .
\end{equation}
Introducing it in Eq.~(\ref{eq:LRUVd4}) and regularizing the UV integral in $Q^2$ with an arbitrary cut-off scale, $\Lambda$,\footnote{Notice that $\lim_{Q^2\to \infty} Q^4 \Pi^{(1+0)}(-Q^2)=0$, together with the analyticity of $\Pi^{(1+0)}(s)$ implies $\int^\infty_0 ds \, s\,\mathrm{Im}\Pi^{(1+0)}(s)=0$ in the chiral limit, also known as a Weinberg Sum Rule \cite{Weinberg:1967kj}. This specifically implies that the integral is independent of $\Lambda$.} leads to 
\begin{equation}
\label{eq:Dasetal}
\Pi_{LR}^{Q}(0)=-\frac{3}{(4\pi)^2\pi} \int^{\infty}_0 \, ds\,  s\,  \log\frac{s}{\Lambda^2} \,\mathrm{Im}\Pi^{(1+0)}(s) \, ,
\end{equation}

In the chiral expansion one only has a contribution at $\mathcal{O}(p^0)$ from Eq.~(\ref{eq:lageffzero}), taking $U\supset \mathbb{I}$,
\begin{equation}
\Pi_{LR}^{Q}(0)=2C \, .
\end{equation}
One can then identify both realizations of $\Pi_{LR}^{Q}(0)$ to find, in the chiral limit
\begin{equation}\label{eq:moussallampionsumrule}
\frac{F^2\Delta m_{\pi^+}^2}{4\pi\alpha}=2C=-\frac{3}{(4\pi)^2\pi} \int^{\infty}_0 \, ds\,  s\,  \log\frac{s}{\Lambda^2} \,\mathrm{Im}\Pi^{(1+0)}(s) \, ,
\end{equation}
which is the pion sum rule as derived in Ref.~\cite{Moussallam:1997xx}.

\section{The finite-energy pion sum rule in the chiral limit}
\label{sec:pionsrchfin}

Experimental input is only available up to a finite energy, so we need to introduce a cutoff in the integral at a scale $s_0$, which in practice we take to be of order the $\tau$-mass scale. We then rewrite the pion sum rule as a finite-energy sum rule, still in the chiral limit. In this limit, the longitudinal contribution vanishes, $Q^2\Pi^{(0)}(-Q^2)=0$. We also simplify the notation by defining
\[
\Pi(-Q^2)\equiv \Pi^{(1+0)}(-Q^2)= \Pi^{3,(1+0)}_{V-A}(-Q^2)\, .
\]
Finally, we split the integral in Eq.~(\ref{eq:LRUVd4}) into two pieces,
\begin{equation}\label{eq:splitting}
-\frac{(4\pi)^2}{3s_0^2}\Pi_{LR}^{Q}(0)=\int^{\infty}_0 \frac{dQ^2}{s_0} \, \frac{Q^2}{s_0}\Pi(-Q^2)=
\underbrace{\int_{s_0}^{\infty}\frac{dQ^2}{s_0} \, \frac{Q^2}{s_0}\Pi(-Q^2)}_{\equiv \hat{\Pi}^{>}_{LR}(0,s_0)}
+ \underbrace{\int_{0}^{s_0}\frac{dQ^2}{s_0} \, \frac{Q^2}{s_0}\Pi(-Q^2)}_{\equiv \hat{\Pi}^{<}_{LR}(0,s_0)} \, .
\end{equation}

\begin{figure}[tb]
\centering
\includegraphics[width=0.5\linewidth]{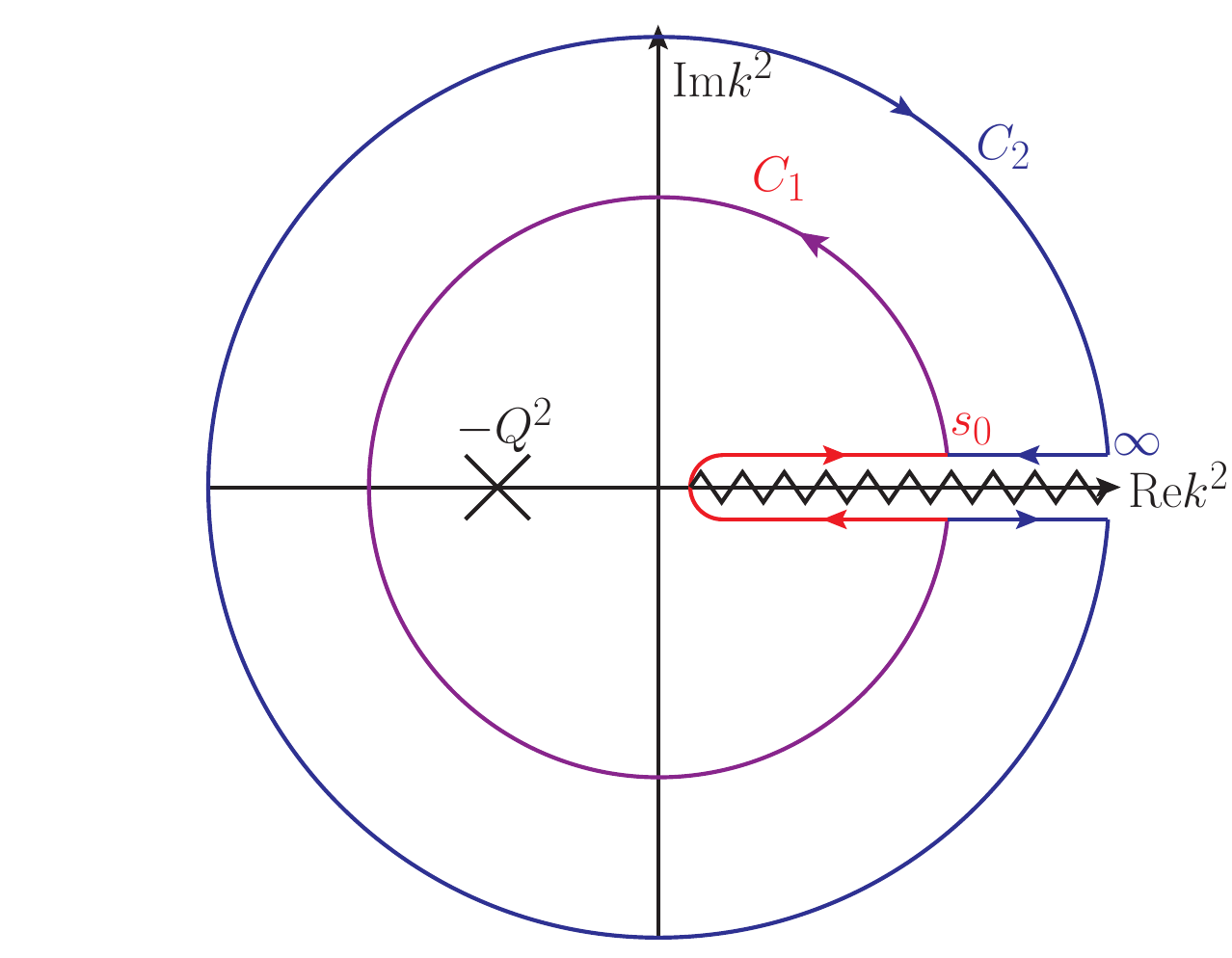}
\caption{Integration contours in the complex $k^2$ plane used to derive the finite-energy representation of $\Pi(-Q^2)$. The jagged line denotes the physical cut along the positive real axis, split at $s_0$; the cross marks the pole at $k^2=-Q^2$ enclosed by $\mathcal{C}_1(s_0)$, while $\mathcal{C}_2(s_0)$ denotes the contour used in the treatment of duality-violating contributions.}
\label{fig:circuit}
\end{figure}

For sufficiently large $s_0$ one can use the OPE of the correlator~\cite{Shifman:1978bx} for the first Euclidean integral. The second term involves $\Pi(-Q^2)$ at $Q^2\sim \Lambda_{\mathrm{QCD}}^2$, and is therefore nonperturbative; it might be feasible to compute it in the chiral limit on the lattice. Here we instead rely on analyticity in the complex $k^2$ plane. To this end we define two integration contours $\mathcal{C}_{1}(s_0)$ and $\mathcal{C}_2(s_0)$ with $k^2 \in \mathbb{C}$ according to
\begin{align}
    \mathcal{C}_1 (s_0) & = \left\{  |k|^2 = s_0 ,\, \arg(k^2)\in [\delta , 2\pi-\delta ] \right\} \, \cup 
    \, \left\{ k^2= s + i \epsilon , \, s\in [s_{th}, s_0] \right\}
    \nonumber \\
    & \, \cup \, 
    \left\{ k^2= s - i \epsilon , \, s\in [s_0,s_{th}] \right\}
    \equiv
    \partial \mathcal{C}_{1} (s_0)\, \cup \, \mathcal{C}_{1}^{\textrm{cut}}(s_0, s_{th}) \, ,
    \\
    \mathcal{C}_2 (s_0) & = \left\{  |k|^2 = \infty ,\, \arg(k^2)\in [ 2\pi-\delta , \delta ] \right\} \, \cup \,
    \left\{ k^2= s + i \epsilon , \, s\in [\infty, s_{0}] \right\}
    \nonumber
    \\
    &
    \, \cup \, 
    \partial \mathcal{C}_{1} (s_0)
    \, \cup \, 
    \left\{ k^2= s - i \epsilon , \, s\in [s_{0}, \infty] \right\}
    \equiv\partial \mathcal{C}_{2} (\infty )\, \cup \,  \partial \mathcal{C}_{1} (s_0) \, \cup \, \mathcal{C}_{2}^{\textrm{cut}}(s_0, \infty ) \, .
\end{align}
Here $\delta, \epsilon >0$ are introduced to avoid the physical cut starting at $s_{th}$ on the real line, and the contours are shown in Fig.~\ref{fig:circuit}. It should be noted that $\mathcal{C}_{2} (s_0)$ is negatively oriented. Inside the contour $\mathcal{C}_{1}(s_0)$, the function $\frac{k^2}{k^2+Q^2}\Pi(k^2)$ is analytic except at the pole $k^2 = -Q^2$, so that the integral
\begin{equation}
\frac{1}{2\pi i}\int_{\mathcal{C}_1(s_0)} \frac{dk^2}{s_0} \frac{k^2}{k^2+Q^2} \Pi(k^2) =-\frac{Q^2}{s_0} \, \Pi(-Q^2) \, .
\end{equation}
The right-hand side is precisely the sought integrand in Eq.~(\ref{eq:splitting}) up to an overall factor.
We can split the contour integral on the left-hand side into its two pieces according to
\begin{equation}
\begin{aligned}
\frac{Q^2}{s_0} \, \Pi(-Q^2)
&
= 
-\frac{1}{2\pi i}
\left(
\int_{\mathcal{C}_{1}^{\textrm{cut}}(s_0, s_{th})}
+
\int_{\partial\mathcal{C}_{1}(s_0)}  
\right) 
\frac{dk^2}{s_0} \frac{k^2}{k^2+Q^2} \Pi(k^2)\, .
\end{aligned}
\end{equation}
From the Schwarz reflection principle we replace the integral around the cut with the imaginary part of $\Pi$ evaluated on the real line, 
\begin{equation}
\begin{aligned}
\frac{Q^2}{s_0} \, \Pi(-Q^2)
&=-\frac{1}{\pi}\int_{s_{th}}^{s_0}\frac{dk^2}{s_0}\frac{k^2}{k^2+Q^2} \mathrm{Im}\,\Pi(k^2) 
-\frac{1}{2\pi i}\int_{\partial\mathcal{C}_{1}(s_0)} \frac{dk^2}{s_0} \frac{k^2}{k^2+Q^2} \Pi(k^2)\, .
\end{aligned}
\end{equation}
For the integral over $\partial \mathcal{C}_1(s_0)$ we will make use of the asymptotic behaviour of the quark-hadron duality violating (DV) piece $\Pi^{\mathrm{DV}}\equiv \Pi-\Pi^{\mathrm{OPE}}$~\cite{Chibisov:1996wf,Cata:2008ye,Gonzalez-Alonso:2010kpl,Boito:2017cnp}, and the analyticity of both $\Pi$ and $\Pi^{\mathrm{OPE}}$. We first write
\begin{equation}
\begin{aligned}
\frac{Q^2}{s_0} \, \Pi(-Q^2)
&=-\frac{1}{\pi}\int_{s_{th}}^{s_0}\frac{dk^2}{s_0}\frac{k^2}{k^2+Q^2} \mathrm{Im}\,\Pi(k^2) 
-\frac{1}{2\pi i}\int_{\partial\mathcal{C}_{1}(s_0)} \frac{dk^2}{s_0} \frac{k^2}{k^2+Q^2} \Pi ^{\textrm{OPE}}(k^2)
 \\
& -\frac{1}{2\pi i}\int_{\partial\mathcal{C}_{1}(s_0)} \frac{dk^2}{s_0} \frac{k^2}{k^2+Q^2} \Pi ^{\textrm{DV}}(k^2)
\, .
\end{aligned}
\end{equation}
The integral on the second line can be replaced by utilizing the analyticity inside the contour $\mathcal{C}_2(s_0)$ in Fig.~\ref{fig:circuit}, i.e.
\begin{equation}
\begin{aligned}
0
&=
-\frac{1}{2\pi i}\int_{\mathcal{C}_2(s_0)} \frac{dk^2}{s_0} \frac{k^2}{k^2+Q^2} \Pi^{\mathrm{DV}}(k^2)
\\
&=-\frac{1}{2\pi i}\int_{\partial \mathcal{C}_{1}(s_0)} \frac{dk^2}{s_0} \frac{k^2}{k^2+Q^2} \Pi^{\mathrm{DV}}(k^2)
-\frac{1}{2\pi i}\int_{\partial \mathcal{C}_{2}(\infty)} \frac{dk^2}{s_0} \frac{k^2}{k^2+Q^2} \Pi^{\mathrm{DV}}(k^2)
\\
&
+\frac{1}{\pi}\int_{s_{0}}^{\infty}\frac{dk^2}{s_0}\frac{k^2}{k^2+Q^2} \mathrm{Im}\,\Pi^{\mathrm{DV}}(k^2) 
\, .
\end{aligned}
\end{equation}
Using that the contribution of the duality-violating part to $\partial \mathcal{C}_{2}(\infty)$ vanishes this yields
\begin{equation}\label{eq:PiQdisprel}
\begin{aligned}
\frac{Q^2}{s_0} \, \Pi(-Q^2)
&=
-\frac{1}{\pi} \, \int^{s_0}_{s_{th}}\frac{dk^2}{s_0} \frac{k^2}{k^2+Q^2}\, \mathrm{Im}\,\Pi(k^2)
- \frac{1}{2\pi i}\int_{\partial \mathcal{C}_{1}(s_0)} \frac{dk^2}{s_0} \frac{k^2}{k^2+Q^2} \Pi^{\mathrm{OPE}}(k^2)
\\
&\quad -\frac{1}{\pi}\int_{s_{0}}^{\infty} \frac{dk^2}{s_0}\frac{k^2}{k^2+Q^2} \mathrm{Im}\,\Pi^{\mathrm{DV}}(k^2) \, .
\end{aligned}
\end{equation}
Inserting Eq.~(\ref{eq:PiQdisprel}) into the needed integral in Eq.~(\ref{eq:splitting}), we obtain
\begin{equation}
\begin{aligned}
\hat{\Pi}^{<}_{LR}(0,s_0)
&=-\frac{1}{\pi} \int^{s_0}_{s_{th}}\frac{dk^2}{s_0}\mathrm{Im}\,\Pi(k^2) \, \frac{k^2}{s_0}\ln\!\left(\frac{k^2+s_0}{k^2}\right)
\\
&\quad -\frac{1}{2\pi i} \int_0^{s_0} \frac{dQ^2}{s_0} \int_{\partial \mathcal{C}_1(s_0)} \frac{dk^2}{s_0} \frac{k^2}{k^2+Q^2} \Pi^{\mathrm{OPE}}(k^2)
\\
&\quad -\frac{1}{\pi} \int^{\infty}_{s_{0}}\frac{dk^2}{s_0}\mathrm{Im}\,\Pi^{\mathrm{DV}}(k^2) \, \frac{k^2}{s_0}\ln\!\left(\frac{k^2+s_0}{k^2}\right)  \, .
\end{aligned}
\end{equation}
Each dimensional operator in the OPE comes with a series in $\alpha_s$. Truncating each dimension at leading nonzero order, one has~\cite{Shifman:1978bx}
\begin{equation}\label{eq:OPEapprox}
\Pi^{\mathrm{OPE}}(k^2)\approx\sum_{D\geq 6}\frac{\mathcal{O}_D}{(-k^2)^{D/2}} \, ,
\end{equation}
which does not contribute to the contour integral,
\begin{equation}
- \frac{1}{2\pi i}\int_{\partial \mathcal{C}_{1}(s_0)} \frac{dk^2}{s_0} \frac{k^2}{k^2+Q^2} \Pi^{\mathrm{OPE}}(k^2) =0\, . 
\end{equation}
Within this framework, one can consistently add the $\alpha_s$-suppressed logarithmic contributions, but they can be safely neglected at large $s_0$. For the upper Euclidean integral, we may approximate $\Pi(-Q^2)$ by its OPE. Using Eq.~(\ref{eq:OPEapprox}) and the definition of Eq.~(\ref{eq:splitting}) one finds
\begin{equation}
\hat{\Pi}^{>}_{LR}(0,s_0)\approx\sum_{D\geq 6}\frac{2}{D-4}\frac{\mathcal{O}_D}{s_0^{D/2}} \, ,
\end{equation}
and the pion sum rule as a finite-energy sum rule can be written as
\begin{equation}
\begin{aligned}
\frac{-(4\pi)^2}{3s_0^2}\frac{F^2\Delta M_{\pi^+}^2}{4\pi\alpha}
&=-\frac{1}{\pi} \int^{s_0}_{s_{th}}\frac{dk^2}{s_0}\mathrm{Im}\,\Pi(k^2) \, \frac{k^2}{s_0}\ln\!\left(\frac{k^2+s_0}{k^2}\right)
+\sum_{D\geq 6}\frac{2}{D-4}\frac{\mathcal{O}_D}{s_0^{D/2}}
\\
&\quad -\frac{1}{\pi} \int^{\infty}_{s_{0}}\frac{dk^2}{s_0}\mathrm{Im}\,\Pi^{\mathrm{DV}}(k^2) \, \frac{k^2}{s_0}\ln\!\left(\frac{k^2+s_0}{k^2}\right)
+\int_{s_0}^{\infty}\frac{dQ^2}{s_0} \, \frac{Q^2}{s_0}\Pi^{\mathrm{DV}}(-Q^2) \, .
\end{aligned}
\end{equation}

Alternatively, one can eliminate the OPE contribution at this leading level in $\alpha_s$ by integrating $\Pi(k^2)\,\frac{k^2}{s_0}\ln\!\left(1+ \frac{k^2}{s_0}\right)$ along $\mathcal{C}_1(s_0)$, since the logarithm is analytic for $|k^2|<s_0$. One finds
\begin{equation}
\begin{aligned}
\frac{1}{\pi}\int_{s_\mathrm{th}}^{s_0} \frac{dk^2}{s_0}\mathrm{Im}\,\Pi(k^2)\, \frac{k^2}{s_0}\ln\!\left(1+ \frac{k^2}{s_0}\right)
&=-\frac{1}{2\pi i}\int_{\partial \mathcal{C}(s_0)}\frac{dk^2}{s_0}\Pi(k^2) \frac{k^2}{s_0}\ln\!\left(1+ \frac{k^2}{s_0}\right)
\\
&=\sum_{D\geq  6}\frac{2}{D-4}\frac{\mathcal{O}_D}{s_0^{D/2}}
+\int^{\infty}_{s_0} \frac{dQ^2}{s_0}\frac{Q^2}{s_0}\Pi^{\mathrm{DV}}(-Q^2)
\\
&\quad -\frac{1}{\pi}\int_{s_0}^{\infty} \frac{dk^2}{s_0}\mathrm{Im}\,\Pi^{\mathrm{DV}}(k^2)\, \frac{k^2}{s_0}\ln\!\left(1+ \frac{k^2}{s_0}\right) \, .
\end{aligned}
\end{equation}
In the second line we have approximated the contour integral by its OPE and treated the duality-violating contribution by a contour deformation analogous to $\mathcal{C}_2(s_0)$, now also accounting for the logarithmic branch cut. Substituting, one finds
\begin{equation}
\frac{-(4\pi)^2}{3s_0^2}\frac{F^2\Delta M_{\pi^+}^2}{4\pi\alpha}
=\frac{1}{\pi} \int^{s_0}_{s_{th}}\frac{dk^2}{s_0}\mathrm{Im}\,\Pi(k^2) \, \frac{k^2}{s_0}\ln\frac{k^2}{s_0}
+\frac{1}{\pi} \int^{\infty}_{s_{0}}\frac{dk^2}{s_0}\mathrm{Im}\,\Pi^{\mathrm{DV}}(k^2) \, \frac{k^2}{s_0}\ln\frac{k^2}{s_0}\, ,
\end{equation}
as expected: at leading order in $\alpha_s$, which on top of the $\frac{\Lambda_{\mathrm{QCD}}^6}{s_{0}^3}$ power suppression is a good approximation at $s_{0}\sim m_{\tau}^2$, one has $\mathrm{Im}\,\Pi^{\mathrm{OPE}}=0$, so $\mathrm{Im}\,\Pi^{\mathrm{DV}}=\mathrm{Im}\,\Pi$ and one recovers the infinite-energy sum rule of the previous section, Eq.~(\ref{eq:moussallampionsumrule}). 

\section{The finite-energy pion sum rule including quark mass corrections}
\label{sec:pionsrphys}

The splitting of the integral in Eq.~(\ref{eq:splitting}), which determines $\Pi^Q_{LR}(0)$, becomes particularly useful once quark mass corrections are included. As before, the OPE provides an accurate description of the correlators above some scale $s_0$. A new feature then appears: 
$\lim_{Q\to\infty} Q^4\Pi^{(0)}(-Q^2)=0$ does not hold beyond the chiral limit and is already broken at first order in the light-quark masses. Indeed, the corresponding OPE contains a $D=4$ term proportional to $m_q\langle\bar q q\rangle$, which yields a logarithmic divergence.  
This signals that regularization is required, and that additional terms—vanishing in the chiral limit—must be included in the derivation. We therefore re-derive the dispersion relations for $\Pi^Q_{LR}(0)$ including these extra pieces in section~\ref{sec:dispLR}, and obtain the corresponding chiral representation in section~\ref{sec:eftLR}. We then combine both results in section~\ref{sec:combineLR}.

\subsection{Dispersive form of the left-right correlator}
\label{sec:dispLR}
A straightforward generalization of the calculation of the previous section to $d$ dimensions leads to
\begin{equation}
\Pi_{LR}^{Q}(0)=\frac{1}{(4\pi)^{d/2}\Gamma(d/2)}\int_0^{\infty} dQ^2 (Q^2)^{\frac{d-2}{2}}\left[(-d+1)\Pi^{(1+0)}(-Q^2)+(d+\xi-1)\Pi^{(0)}(-Q^2)\right] \, .
\end{equation}

As anticipated, at leading order in $\alpha_s$ the OPE contains a new $D=4$ term linear in the light-quark masses; see e.g.~Ref.~\cite{Braaten:1991qm},
\begin{equation}\label{eq:opeapp}
\Pi^{(0)}(-Q^2)=-\frac{(m_u+m_d) \langle \bar{q}q\rangle}{Q^4} \, .
\end{equation}
This yields
\begin{equation}\begin{aligned}
\Pi_{LR}^{(0)Q,>}(0,s_0)&= -\frac{(m_u+m_d)\langle \bar{q}q\rangle(d+\xi-1)}{(4\pi)^{d/2}\Gamma(d/2)}\int_{s_0}^{\infty} dQ^2 (Q^2)^{\frac{d-6}{2}} \\
&=\frac{(m_u+m_d)\langle \bar{q}q\rangle(d+\xi-1)}{(4\pi)^{d/2}\Gamma(d/2)} (s_0)^{\frac{d-4}{2}}\frac{2}{d-4} \\
&=-\frac{(m_u+m_d)\langle \bar{q}q \rangle }{16\pi^2} \left[(3+\xi)\left(\frac{1}{\hat{\epsilon}} -\ln(s_0)\right)+(1+\xi) \right] \, ,
\end{aligned}
\end{equation}
where we have taken $d=4-2\epsilon$ and defined the {$\overline{\textrm{MS}}$} combination
\begin{equation}\frac{1}{\hat\epsilon}\equiv \frac{1}{\epsilon}-\gamma_E+\ln(4\pi) \, .
\end{equation}
\begin{figure}
    \centering
    \includegraphics[width=0.6\linewidth]{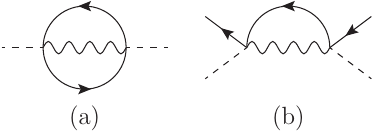}
    \caption{The diagrams at lowest order that lead to counter terms needed. Dashed lines indicate an instertion of the spurion terms of Eq.~(\ref{eq:lagone}), the arrowed lines indicate quarks and the wiggly line a photon. (a) The leading contribution to the $\langle Q_L^2+Q_R^2\rangle$ operator. (b) The type of diagrams contributing to the LR case.}
    \label{fig:counter}
\end{figure}

The spurions $Q_{L;R}$ as defined in Eq.~(\ref{eq:lagone}) do not lead to finite correlators defined by these. This requires adding counterterms. In standard ChPT these lead to the extra high energy terms $H_1$ and $H_2$ at low energies, another example is that in order to get finite vector two-point functions one has to add the operator $P_7$ in the discussion of Ref.~\cite{Gasser:2003hk}. We can start collecting invariant terms that contain the spurions in order to make all correlators finite. Examples of diagrams are shown in  Fig.~\ref{fig:counter}. The one in Fig.~\ref{fig:counter}(a) only starts at two-loop order. Other counter terms can be derived from graphs like the one shown in Fig.~\ref{fig:counter}(b). The relevant one for our calculation is
\begin{equation}
\mathcal{L}\supset -Z_s \, \bar{q}_R\, (Q_R\mathcal{M} Q_L)\, q_L -Z_s \, \bar{q}_L\, (Q_L\mathcal{M}^{\dagger} Q_R)\, q_R \, ,
\end{equation} 
We use the $\overline{\textrm{MS}}$ prescription to determine the counterterms.  The calculation for the one relevant here leads to
\begin{equation}
Z_s=Z_s (\mu _{\textrm{SD}})=(-\xi-3)\frac{\mu _{\textrm{SD}}^{-2\epsilon}}{16\pi^2\hat{\epsilon}} \, .
\end{equation}
As can be seen, the counterterm depends on the choice of gauge and the result agrees with Refs.~\cite{Bijnens:1996kk,Moussallam:1997xx}. Once all the spurions are saturated back to $Q_{L}=Q_{R}=\frac{e}{3}\,\mathrm{diag}(2,-1,-1)$, this set of counterterms reduces to the usual QED ones and one recovers the usual renormalized theory.

The corresponding contribution\footnote{The operator contributes to our correlators since we have to allow for the nonperturbative vacuum expectation value.} is
\begin{equation}\begin{aligned}
\Pi_{LR,\mathrm{ct}}^{Q}(0)
&=-(m_u+m_d)\langle \bar{q}q \rangle Z_s=(m_u+m_d)\langle \bar{q}q \rangle (\xi+3)\frac{\mu_{\textrm{SD}}^{-2\epsilon}}{16\pi^2\hat{\epsilon}} \, .
\end{aligned}
\end{equation}
We thus find
\begin{equation}\label{eq:short}
\Pi_{LR}^{(0)Q,>}(0,s_0)+\Pi_{LR,ct}^Q(0)=-\frac{(m_u+m_d)\langle \bar{q}q \rangle }{16\pi^2} \left[1+\xi-(3+\xi)\ln\frac{s_0}{\mu_{\textrm{SD}}^2}\right] \, .
\end{equation}

At the working order, the same discussion as in the previous section applies to the
$\Pi^{(1+0)}$ correlator, except that, in addition to the hadronic cut, one must account for
the fact that the pion-pole singularity does not exactly cancel against the kinematic weight,
$q^2\Pi^{(1+0)}(q^2)=q^2\,\frac{F_\pi^2}{q^2-m_\pi^2}$.
One can absorb this either by shifting the lower limit to
$\hat{s}_{th}=m_{\pi}^{2}-\epsilon$, i.e. including
$\mathrm{Im}\Pi^{(1+0)}\supset -\pi\, F_{\pi}^2 \,\delta(k^2-m_{\pi}^2)$,
or, equivalently, by keeping the continuum threshold at $s_{th}=4 m_{\pi}^{2}$ and treating
the pole as an extra residue in the complex integral. For the $\Pi\equiv\Pi^{(1+0)}$ piece,
this gives, trivially extending the derivation of the previous section,
\begin{equation}
\begin{aligned}
-\frac{16\pi^2}{3s_0^2}\Pi^{Q,(1+0)}_{LR}(0)
&=\frac{1}{\pi} \int^{s_0}_{s_{th}}\frac{dk^2}{s_0}\mathrm{Im}\Pi(k^2) \, \frac{k^2}{s_0}\ln\frac{k^2}{s_0}+\frac{1}{\pi} \int^{\infty}_{s_{0}}\frac{dk^2}{s_0}\mathrm{Im}\Pi^{\mathrm{DV}}(k^2) \, \frac{k^2}{s_0}\ln\frac{k^2}{s_0}\\
&-\frac{F_{\pi}^2m_\pi^2}{s_0^2}\ln\frac{m_\pi^2}{s_0} \, .
\end{aligned}
\end{equation}
Notice $\Pi^{Q,(1+0)}_{LR}(0)$ is, by construction, independent on $s_0$. 

For the remaining longitudinal piece $\Pi_{LR}^{(0)Q,<}(0,s_0)$ we use that, to a very good approximation, the longitudinal part of the correlator is
\begin{equation}
\Pi^{(0)}(q^2)=-F_{\pi}^2\left(\frac{1}{q^2}-\frac{1}{q^2-m_\pi^2} \right) \, .
\end{equation}
We have incorporated the pion pole contribution and used that the contribution of the continuum cut is suppressed by two powers of the light quark masses, dominated by the $\pi(1300)$ resonance, see, e.g.~Ref.~\cite{Bijnens:1994ci}. The kinematic pole at $0$ is such that $q_{\mu}q_\nu\Pi^{\mu\nu}=0$ in the chiral limit. This form also agrees with the one-loop chiral form, see, e.g.~Ref.~\cite{Amoros:1999dp}. One then finds the corresponding contribution
\begin{equation}
\Pi^{<,Q,(0)}_{LR}(0,s_0)=\frac{F_{\pi}^2m_\pi^2(3+\xi)}{16\pi^2}\ln\left(1+\frac{s_0}{m_\pi^2} \right)\approx \frac{F_{\pi}^2m_\pi^2(3+\xi)}{16\pi^2}\ln\left(\frac{s_0}{m_\pi^2} \right) \, ,
\end{equation}
where in the last step we have consistently dropped terms quadratic in the quark masses.

Using that $(m_u+m_d)\langle \bar{q} q\rangle=-F_\pi^2 m_{\pi}^2$ up to higher-order corrections and collecting all pieces, we find
\begin{equation}\label{eq:dispcorr}
\begin{aligned}
16\pi^2\Pi^{Q}_{LR}(0)
&=16\pi^2\left[\Pi^{Q,(1+0)}_{LR}(0)+\Pi^{<,Q,(0)}_{LR}(0,s_0)+\Pi^{>,Q,(0)}_{LR}(0,s_0)+\Pi_{LR,ct}^Q(0)\right]
\\
&=-\frac{3s_0^2}{\pi}\int^{s_0}_{s_{th}}\frac{dk^2}{s_0}\,\mathrm{Im}\Pi(k^2)\,\frac{k^2}{s_0}\ln\frac{k^2}{s_0}
-\frac{3s_0^2}{\pi}\int^{\infty}_{s_{0}}\frac{dk^2}{s_0}\,\mathrm{Im}\Pi^{\mathrm{DV}}(k^2)\,\frac{k^2}{s_0}\ln\frac{k^2}{s_0}
\\
&\qquad +F_{\pi}^2m_\pi^2\left[1+\xi-3\ln\frac{s_0}{\mu_{\mathrm{SD}}^2}+\xi\ln\frac{\mu_{\mathrm{SD}}^2}{m_{\pi}^2} \right] \, .
\end{aligned}
\end{equation}
Therefore $\Pi^{Q,(0)}_{LR}$, which is independent of $s_0$, depends on the gauge choice and on the short-distance scale $\mu_{\mathrm{SD}}$. 

\subsection{The effective left-right correlator}\label{sec:eftLR}
In this section we consider the left-right correlator within an effective field theory framework. A key result is the correlator Eq.~(\ref{eq:chiralrep}), which when combined with
Eq.~(\ref{eq:dispcorr}) yields the final matching condition in Eq.~(\ref{eq:pionsumrulephys}). 

\subsubsection{Chiral Perturbation Theory in the presence of electromagnetism}
Chiral perturbation theory is the low-energy effective field theory of QCD first introduced in Refs.~\cite{Gasser:1983yg,Gasser:1984gg}. For $N_f$ light quarks, this effective field theory is formulated in terms of the $N_f^2-1$ lightest pseudoscalar bosons coupled to external sources.  Inherent to the effective field theory is a systematic expansion in powers of a low-energy momentum, $p^2$. The effective Lagrangian takes the form
\begin{align}\label{eq:chptLag}
    \mathcal{L}_{\chi} = \sum _{n=1}^{\infty} \mathcal{L}_{2n} \, ,
\end{align}
where $\mathcal{L}_{2n}$ is of order $p^{2n}$. Each Lagrangian is given by a sum of monomials, invariant under the same symmetries as QCD, each multiplied by a low-energy constants. The number of low-energy constants (and hence operators) depends on the number of flavors $N_f$. In its most standard form, the Lagrangian of ChPT is known through order $p^8$~\cite{Gasser:1983yg,Gasser:1984gg,PhysRevD.53.315,Bijnens:1999sh,Weber:2008,Bijnens:2018lez,Hermansson-Truedsson:2020rtj}, and to the same order for the odd-intrinsic-parity sector containing the anomaly~\cite{WESS197195,Witten:1983tw,Bijnens:2001bb,Ebertshauser:2001nj,Bijnens:2023hyv}. For a detailed introduction to ChPT see for instance Ref.~\cite{Pich:2018ltt}. 

Beyond the standard formulation, also other low-energy degrees of freedom can be added dynamically. For instance, QED with dynamical photons was introduced in Refs.~\cite{ECKER1989311,Urech:1994hd,Neufeld:1995mu}. The effective Lagrangian for QCD+QED is a combined expansion in $p^2$ and $e^2$. Here we are only interested in $e^2$ corrections in addition to the chiral $p^2$ expansion, and therefore write
\begin{align}
    \mathcal{L}_{\chi}(e^2) 
    & = \sum _{n=1}^{\infty} \mathcal{L}_{2n} (e^2) \\
    \mathcal{L}_{2} (e^2)
    & = \mathcal{L}_{2} + \mathcal{L}_{0}^{e^2}  \, , 
    \\
    \mathcal{L}_{4} (e^2) 
    & = \mathcal{L}_{4} + \mathcal{L}_{2}^{e^2}  \, .
\end{align}
Here $\mathcal{L}_{2n}$ are the ones from Eq.~(\ref{eq:chptLag}). The leading order is thus given by $\mathcal{O}(p^2,p^0 e^2)$, the next-to-leading order $\mathcal{O}(p^4,p^2 e^2)$, and so on. The leading Lagrangian entering here is (for a general gauge parameter $\xi$)~\cite{Urech:1994hd,Bijnens:2006mk,Agadjanov:2013lra}
\begin{align}\label{eq:LO}
\mathcal{L}_{2}(e^2) = -\frac{1}{4}\, F_{\mu \nu}F^{\mu \nu} - \frac{1}{2 \xi} \, (\partial _\mu A^\mu)^2 +
\frac{F_0^{2}}{4}\langle u_{\mu}u^{\mu}+\chi _{+}\rangle 
+  C\, \langle \mathcal{Q}_L \mathcal{Q}_R\rangle
\, .
\end{align}
The above Lagrangian is in the basis with $u$ such that $u^2 = U$ in the original formulation of ChPT~\cite{Gasser:1983yg,Gasser:1984gg}. For details about the notation see e.g.~Ref.~\cite{Bijnens:2006mk}. The low-energy constant $C$ is directly related to the electromagnetic shift to the pion and kaon masses. Working in the limit $m_u - m_d = 0$ and writing $ \hat{m} = (m_u+m_d)/2 $ one has
\begin{align}
m_{\pi^{\pm}}^2 
& \stackrel{\textrm{LO}}{=}
2 B_0 \, \hat{m} +  \frac{2C\, e^2}{F_0^2} =  m_{\pi}^2 +  \frac{2C\, e^2}{F_0^2} \, ,
\nonumber \\
m_{K^{\pm}}^2 
& \stackrel{\textrm{LO}}{=}
B_0 \left(\hat{m}+m_s\right) +  \frac{2C\, e^2}{F_0^2} = m_{K}^2 +  \frac{2C\, e^2}{F_0^2} 
\, ,
\end{align}
where 
$B_0 $ is a low-energy constant related to the quark condensate. 
The above relation can be obtained by putting $\mathcal{Q}_L = u Q_L u ^\dagger $, $\mathcal{Q}_R = u^\dagger Q_R u  $ with $Q_L = Q_R = e\, \textrm{diag}(2,-1,-1)/3$. Below we will saturate $Q_L$ and $Q_R$ in two ways and extract a 2-point left-right correlator. 

For the pions at next-to-leading order we need the NLO Lagrangian $ \mathcal{L}_{4} (e^2)$~\cite{Urech:1994hd,Agadjanov:2013lra}, from which the masses are found to be
\begin{equation}
\begin{aligned}
m_{\pi^\pm}^2  
&
\stackrel{\textrm{NLO}}{=} 
 m_{\pi}^2+ 
\frac{2C\, e^2}{F_0^{2}}
- e^2 \frac{1}{16\pi^{2}} m_\pi^{2}\!\left( 3\ln\frac{m_\pi^{2}}{\nu_\chi^{2}} - 4 \right) \\
&\quad 
- e^2 \frac{1}{8\pi^{2}} \frac{C}{F_0^{4}}
\left[ 2 m_\pi^{2}\, \ln\frac{m_\pi^{2}}{\nu_\chi^{2}} 
+ m_K^{2}\ln\frac{m_K^{2}}{\nu_\chi^{2}} \right] \\
&\quad - 16 e^2 \frac{C}{F_0^{4}}
\left[ \left(m_\pi^{2} + 2 m_K^{2}\right) L_4^r + m_\pi^{2} L_5^r \right] \\
&\quad - \frac{4}{9} \, e^2 m_\pi^{2} \left[ 6K_1^r+6 K_2^r+5 K_5^r + 5 K_6^r -6 K_7^r - 15 K_8 ^r -5 (K_9^r+K_{10}^r) - 18( K_{10}^r + K_{11}^r)\right] \\
&\quad + 8 e^2 m_K^{2} K_8^r \, ,
\end{aligned}
\end{equation}
\begin{equation}
\begin{aligned}
m_{\pi^0}^2  
&
\stackrel{\textrm{NLO}}{=} 
 m_{\pi}^2 
+ e^2 \frac{1}{8\pi^{2}} \frac{C}{F_0^4}\, m_\pi^{2}\!\left( \ln\frac{m_\pi^{2}}{\nu_\chi^{2}} +1 \right)  - \frac{2}{9}\,  e^2 m_\pi^{2} \Big[ 12K_1^r + 12K_2^r -18 K_3^r 
\\
&\quad
+ 9K_4^r+10 K_5^r + 10 K_6^r-12 K_7^r -12 K_8^r -10 (K_9^r+ K_{10}^r)\Big] \, .
\end{aligned}
\end{equation}
The $K_i^r$ are renormalized low-energy constants depending on the chiral renormalisation scale $\nu _{\chi}$. They are related to the bare $K_i$ $\overline{\textrm{MS}}$ (with a conventional +1 for chiral perturbation theory) through
\begin{align}
    K_i & = K_i^{r}(\nu _\chi) + \Sigma _i \, \lambda \, ,
    \\
    \lambda  & = \frac{\nu _\chi ^{d-4}}{16\pi^2} \left\{ \frac{1}{d-4} - \frac{1}{2}\left(-\gamma_E+\log 4\pi +1\right) \right\} \, .
\end{align}
The constants $\Sigma _i$ are given in a general $R_\xi$ gauge in Eq.~(5.4) of Ref.~\cite{Agadjanov:2013lra}. It is only $K_{9,\ldots,13}^r$ which are dependent on the gauge parameter $\xi$~\cite{Bijnens:1996kk,Moussallam:1997xx,Agadjanov:2013lra,Bijnens:2014lea}. The constants $K_{9,\ldots,12}^r$ in addition depend on the QCD renormalisation scale $\mu _{\textrm{SD}}$~\cite{Bijnens:1996kk,Moussallam:1997xx,Agadjanov:2013lra,Bijnens:2014lea}. Certain combinations of $K_i$ such as $K_{10}^r+K_{11}^r$ appearing above are independent of both the gauge parameter $\xi$ and the QCD renormalisation scale $\mu _{\textrm{SD}}$~\cite{Agadjanov:2013lra,Bijnens:2014lea}. However, the combination $K_{9}+K_{10}$ appearing in $m_{\pi^\pm}^2$ is only independent of $\xi$ but not of $\mu _{\textrm{SD}}$~\cite{Moussallam:1997xx}. This is related to the ambiguity of separation-scheme dependence in QCD+QED~\cite{Bijnens:1996kk,Moussallam:1997xx,Gasser:2003hk}. These subtleties are essential for reliable high-precision determinations of hadronic observables~\cite{Aoyama:2020ynm,FlavourLatticeAveragingGroupFLAG:2024oxs,Aliberti:2025beg}. 

Using the above expressions for the pion masses we can form the difference $\Delta m_\pi^2 = m_{\pi^\pm}^2-m_{\pi^0}^2$. The result is~\cite{Urech:1994hd}
\begin{equation}
\begin{aligned}
\Delta m_\pi^2
&\stackrel{\textrm{NLO}}{=} m_{\pi^+}^2 - m_{\pi^0}^2 \\
&= \frac{2 e^2 C}{F_0^{2}}
- e^2 \frac{1}{16\pi^{2}} \, m_\pi^{2}\!\left( 3\ln\frac{m_\pi^{2}}{\nu_\chi^{2}} - 4 \right) \\
&\quad - e^2 \frac{1}{8\pi^{2}} \frac{C}{F_0^{4}}
\left[ m_\pi^{2}\!\left( 3\ln\frac{m_\pi^{2}}{\nu_\chi^{2}} + 1 \right)
+ m_K^{2}\ln\frac{m_K^{2}}{\nu_\chi^{2}} \right] \\
&\quad - 16 e^2 \frac{C}{F_0^{4}}
\left[ \left(m_\pi^{2} + 2 m_K^{2}\right) L_4^r + m_\pi^{2} L_5^r \right] \\
&\quad + 2 e^2 m_\pi^{2} \left[ -2K_3^r + K_4^r + 2K_8^r + 4(K_{10}^r + K_{11}^r) \right] \\
&\quad + 8 e^2 m_K^{2} K_8^r \, .
\end{aligned}
\end{equation}
The pion mass shift is independent of  $\mu_{\textrm{SD}}$ and $\xi$ in $K_{10}^r$ and $K_{11}^r$, since only the invariant combination $K_{10}^r+K_{11}^r$ appears on the right-hand side.
Next we wish to rewrite $C$ in terms of $\Delta m_\pi^2$ by perturbatively solving the above equation. We further insert the physical pion decay constant $F_\pi$ instead of $F_0$. For the latter we have
\begin{align}
   F_0 \stackrel{\textrm{NLO}}{=} F_\pi (1-F_{4} ) \, , 
\end{align}
where $F_4$ is the NLO correction. It should be noted that $F_4$ can be taken in the isospin-symmetric limit here since corrections would be beyond the order studied. The result is~\cite{Gasser:1983yg,Gasser:1984gg}
\begin{align}
F_4
= 
-\frac{1}{32\pi^{2}F_0^{2}}
\Big[\,2m_\pi^{2}\log\!\frac{m_\pi^{2}}{\nu_\chi ^{2}}
+ m_K^{2}\log\!\frac{m_K^{2}}{\nu_\chi^{2}} \Big]
+ \frac{4m_\pi^{2}}{F_0^{2}}\,L_{5}^r
+ \frac{8m_K^{2}+4m_\pi^{2}}{F_0^{2}}\,L_{4}^r
\, .
\end{align}
Combining everything, we find the NLO relation
\begin{align}\label{eq:CNLO}
C 
&   \stackrel{\textrm{NLO}}{=} 
  \frac{\Delta m_\pi^2}{2e^2} \left[ F_\pi^2+\frac{m_\pi^2}{16 \pi ^2   }   \right] -\frac{m_\pi^2\, F_{\pi}^2}{8 \pi ^2 }  
  \nonumber \\
  &
  +F_{\pi}^2  \Bigg\{ 
    -4 K_{8}^r
   m_K^2
   -4 m_\pi^2 \left(
   K_{10}^r+ K_{11}^r-\frac{1}{2} K_{3}^r+\frac{1}{4} K_{4}^r+\frac{1}{2}  K_{8}^r\right)
   \nonumber \\
    &
    +\frac{1}{16 \pi ^2}\, \log \frac{m_K^2}{\nu _{\chi}^2} \, m_K^2 \left[
     \frac{\Delta m_\pi^2}{e^2 F_\pi^2 }
   \right]
   + \frac{1}{32 \pi ^2}\log \frac{m_\pi^2}{\nu _{\chi}^2}\, m_\pi^2 \left[ 3 + 5\,  \frac{\Delta m_\pi^2}{e^2 F_\pi^2 }\right] 
   \Bigg\}
\end{align}
As can be seen, the dependence on the LECs $L_{4,5}^r$ has cancelled from the change to $F_\pi$. We will use this relation for $C$ below.

\subsubsection{The left-right correlator in chiral perturbation theory}
\begin{figure}
    \begin{center}
\includegraphics[width=0.4\linewidth]{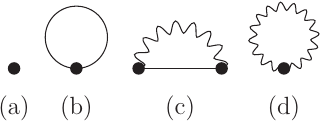}
    \end{center}
     \caption{The diagrams contributing to $\Pi ^{Q,\chi}_{\textrm{LR}}(q^2)$ through NLO in the chiral counting.}
     \label{fig:diagrams} 
    \end{figure}

Next we consider the left-right correlator within chiral perturbation theory. For the order of interest here we need the Lagrangians $\mathcal{L}_2(e^2)$ and $\mathcal{L}_4(e^2)$ above. The correlator is given by
\begin{align}\label{eq:chiralLRcorr}
\Pi ^{Q,\chi}_{\textrm{LR}}(q^2)= \frac{1}{e^2} \int d^4x \,  e^{i q\cdot x} \,   \langle 0\vert T\{\phi_L(x)\phi_R(0)\} \vert 0\rangle \, ,
\end{align}
where $\phi _{L,R}$ have been introduced into the Lagrangian through the replacement $Q_{L,R}\rightarrow \phi _{L,R}Q_{L,R}$. 
We then saturate the matrices $Q_L = Q_R = Q$ with\footnote{The $1/e^2$ in Eq.~(\ref{eq:chiralLRcorr}) is introduced to remove the $e$ in the definition in $Q$. }
\begin{align}
    Q & = e\, \textrm{diag}(1,-1,0) \, . 
\end{align}
The diagrams through NLO are shown in Fig.~\ref{fig:diagrams}(a)--(d). Since we are working without a photon mass, diagram (d) is zero in the dimensional regularization adopted here. 
From diagram (a) we obtain 
\begin{align}
    \Pi ^{Q,\chi}_{\textrm{LR}}(q^2) 
    \stackrel{(a)}{=} 
     2C
    + \, F_0^2\left[ m_\pi^2 \left(4K_8+8K_{10}\right)+m_K^2 \, 8K_8+ q^2 2 K_{13}\right]\, . 
\end{align}
The first term comes from the operator $ C\langle \mathcal{Q}_L \mathcal{Q_R}\rangle $ in $\mathcal{L}_2(e^2)$. The second one comes from $\mathcal{L}_4(e^2)$.
Note that the $K_i$ appearing here still are unrenormalized.
Their divergences will eventually cancel against the loop diagrams (b) and (c). For the loop diagrams the standard tadpole and bubble integrals appear,
\begin{align}
A(m^2) &= \frac{m^2}{16\pi^2}\left(\frac{1}{\bar\epsilon}-\log m^2 \right) \, ,\\
B(m^2,0,q^2) &= \frac{1}{16\pi^2}\left(\frac{1}{\bar\epsilon}-\log m^2 +1+\left(1-\frac{m^2}{q^2}\right)\log\left(\frac{m^2}{m^2-q^2}\right)\right) \, , \\
\hat B(m^2,q^2) &= \frac{1}{16\pi^2}\left(-2\frac{m^2}{q^2}\log\left(\frac{m^2}{m^2-q^2}\right)+2\right)\, .
\end{align}
Since our massless photon propagator in an $R_\xi$ gauge explicitly depends on $\xi$ we also need the loop function\footnote{We note that with a photon mass regularization the propagator takes the form $-i/(k^2-m_\gamma ^2) \left[g_{\mu\nu}+(\xi-1)\, k_\mu k_\nu / (k^2-\xi \, m_\gamma ^2)\right]$~\cite{Moussallam:1997xx}, so that $\xi$ no longer is an overall factor in loop integrals. This means that the integrals results presented here are strictly for the $m_\gamma =0$ case, and would contain additional dependence on $\xi$ if a photon mass were used. }
\begin{align}
F_Q(m^2,q^2) = m^2 B(m^2,0,q^2)+\frac{\left(\xi-1\right)}{4}\left[A(m^2)+\left(m^2-q^2\right)\left(3 B(m^2,0,q^2)+\hat B(m^2,q^2)\right)\right] \, . 
\end{align}
For Feynman gauge $\xi = 1$ the whole second term drops out. In diagram Fig.~\ref{fig:diagrams}(b) the particles in the loop are the charged pion as well as both charged and neutral kaons. This gives
\begin{align}
 \Pi ^{Q,\chi}_{\textrm{LR}}(q^2) 
    \stackrel{(b)}{=}  \frac{2 C}{F^2_0}\left[4 A(m_\pi^2)+2A(m_K^2)\right]\,.
\end{align}
Fig.~\ref{fig:diagrams}(c) has the neutral pion in the loop together with a photon, and yields
\begin{align}
\Pi ^{Q,\chi}_{\textrm{LR}}(q^2) 
    \stackrel{(c)}{=}  F_0^2\, F_Q(m_\pi^2,q^2) \, . 
\end{align}
Adding all pieces together at $q^2 = 0$ we obtain the finite correlator
\begin{align}
     \Pi ^{\chi}_{\textrm{LR}}(0) 
    & 
    = 
       2\, C  
      +  m_K^2 \left[
           8\, F_0^2\, K^r_8
          - \frac{4\, C}{16\pi^2 \, F_0^2 }\, \log \frac{m_K^2}{\nu _\chi ^2} 
        \right]
        \nonumber \\
       & 
       +  m_{\pi}^2  
       \left[
           8\, F_0^2 \, K^r_{10}
          + 4\, F_0^2 \, K^r_8
          - \frac{8\, C}{16\pi^2\, F_0^2}\, \log \frac{m_\pi^2}{\nu _\chi ^2}
          - \frac{ F_0^2\, \xi}{16\pi^2}\,\log \frac{m_\pi ^2}{\nu _\chi ^2} 
        \right] \, . 
\end{align}
Here we see that the left-right correlator is explicitly $\xi$ dependent, and we stress that this relation holds true for general $\xi$ and $m_\gamma = 0$. The IR divergence has in our calculation been regulated using dimensional regularization. Next we rewrite this equation in terms of $F_\pi$,
\begin{equation}
\begin{aligned}
\Pi_{LR}^{Q,\chi }(0)&=  F_\pi^{2}\!\left[ 8 K_{8}^{r}\, m_{K}^{2} + (8 K_{10}^{r} + 4 K_{8}^{r})\, m_{\pi}^{2} \right]
-  \frac{F_{\pi}^{2} m_{\pi}^{2}}{16\pi^{2}} \,  \left( \xi\, \log\frac{m_{\pi}^{2}}{\nu_{\chi}^{2}} \right) 
\\
&
+  2 C \left[ 1 - \frac{1}{F_{\pi}^{2}} \left( \frac{m_{K}^{2}}{8\pi^{2}} \log\frac{m_{K}^{2}}{\nu_{\chi}^{2}} + \frac{m_{\pi}^{2}}{4\pi^{2}} \log\frac{m_{\pi}^{2}}{\nu_{\chi}^{2}} \right) \right] \, .
\end{aligned}
\end{equation}
Further replacing $C$ with the expression from Eq.~(\ref{eq:CNLO}) we obtain
\begin{align}\label{eq:chiralrep}
  \Pi_{LR}^{Q,\chi }(0) 
  &
  =  \frac{\Delta m_{\pi}^2}{e^2} \left( F_\pi^2+\frac{m_\pi^2}{16\pi^2}\right)
  +   2F_\pi^2\, m_\pi^2 \left(
  -4 
    K_{11}^r
   +2 K_{3}^r
   - K_{4}^r
   \right)
   \nonumber \\
   &
    -\frac{ F_\pi^2\, m_\pi^2 }{4 \pi
   ^2} 
   +  \frac{ m_\pi^2}{16\pi^2}  \Bigg( 
   (3-\xi ) 
   F_\pi^2 
   +\frac{\Delta m_{\pi}^2
  }{e^2}  
  \Bigg) \log \frac{m_\pi^2}{\nu _\chi ^2} \, . 
\end{align}
Here we note that all previous dependence on kaon masses has canceled. 
In the expression above we see a dependence on $K_{3}^r$, $K_{4}^r$ and $K_{11}^r$ through the linear combination $Y=-4 K_{11}^r+2 K_{3}^r- K_{4}^r$. Note that $K_{11}^r$ is both gauge and QCD-scale dependent. Using numerical values for the LECs from Ref.~\cite{Bijnens:2014lea}, which are without uncertainties, one finds 
\begin{align}\label{eq:Yleccomb}
    Y = -4 K_{11}^r+2 K_{3}^r- K_{4}^r 
    =  -1.2\times 10^{-3}\,  \quad  \big[\xi = 1, \,\,  \mu_{\textrm{SD}}= 1 \, \textrm{GeV}, \, \, \nu_{\chi}= 0.77 \, \textrm{GeV}\big] \, .
\end{align}
We may further investigate the dependence on renormalization scales and $\xi$ in $Y$. To this end we equate the chiral representation of the left-right correlator Eq.~(\ref{eq:chiralrep}) with the dispersive one in Eq.~(\ref{eq:dispcorr}) at two sets of scales and gauge parameters. This yields 
\begin{align}\label{eq:Yrun}
Y(\mu_{\mathrm{SD}},\xi,\nu_{\chi})
& =
Y(\mu_{\mathrm{SD}}^{0},\xi^{0},\nu_{\chi}^{0})
+\frac{1}{32\pi^{2}}
\Bigg[
(3+\xi)\ln\frac{\mu_{\mathrm{SD}}^{2}}{\mu_{\mathrm{SD}}^{0\,2}}
+\left(3-\xi+\frac{\Delta m_\pi^{2}}{e^{2}F_\pi^{2}}\right)\ln\frac{\nu_{\chi}^{2}}{\nu_{\chi}^{0\,2}}
\nonumber \\
&
\qquad
\qquad
\qquad
\qquad
\qquad
+(\xi-\xi^{0})\left(1+\ln\frac{\mu_{\mathrm{SD}}^{0\,2}}{\nu_{\chi}^{0\,2}} \right) 
\Bigg]  \, .
\end{align}
This scaling is consistent with the results of Ref.~\cite{Moussallam:1997xx} and the chiral renormalization-scale dependence in Refs.~\cite{Urech:1994hd,Agadjanov:2013lra}.

Considering that the exact choice of $\mu_{\mathrm{SD}}$ and $\nu_{\chi}$ for those estimates are relatively loosely defined, we estimate uncertainties with the standard criteria of adding quadratically a variation of the renormalization scales, $\mu_{\mathrm{SD}}\in [0.9,1.2]$ GeV and $\nu_{\mathrm{\chi}}\in [0.6,1]$ GeV, see e.g.~Ref.~\cite{Cirigliano:2019cpi}. This gives
\begin{align}
32\pi^2 Y
& =
-0.4 \pm 2.2 \quad  \big[\xi = 1, \,\,  \mu_{\textrm{SD}}= 1 \, \textrm{GeV},\, \, \nu_{\chi}= 0.77 \, \textrm{GeV}\big] 
\nonumber \\
& 
\approx 
32\pi^2 \, (-1.3\pm 6.9)\times 10^{-3}
\, . \label{eq:Kestimates}
\end{align}

\subsection{Matching representations}\label{sec:combineLR}
Combining Eqs.~(\ref{eq:dispcorr}) and~(\ref{eq:chiralrep}) one finds 
\begin{equation}\label{eq:pionsumrulephys}
\begin{aligned}
&\frac{8\pi}{\alpha}\frac{\Delta m_{\pi}}{m_\pi}
+32\pi^2\left(-4K_{11}^r+2K_{3}^r-K_{4}^r\right)
-5-\xi
+3\ln\frac{m_\pi^2}{\nu_\chi^2}
-\xi\ln\frac{\mu_{\mathrm{SD}}^2}{\nu_\chi^2}
+3\ln\frac{s_0}{\mu_{\mathrm{SD}}^2}
\\
&=-\frac{3}{ F_{\pi}^2m_{\pi}^2}\left(
\frac{1}{\pi}\int_{s_{th}}^{s_0} dk^2\,\mathrm{Im}\Pi(k^2)\, k^2\ln\frac{k^2}{s_0}
+ \delta_{\mathrm{DV}}(s_0)\right)
\left[1-\frac{m_\pi^2}{(4\pi F_{\pi})^2}\left(1+\ln\frac{m_\pi^2}{\nu_\chi^2}\right)\right]
\, .
\end{aligned}
\end{equation}
which is independent of $\xi$, $s_0$, $\nu_\chi$ and $\mu_{\mathrm{SD}}$, up to higher-order chiral corrections, and generalizes the pion sum rule by including all corrections linear in the quark masses. In a more compact form one has\footnote{For numerical reference, $\frac{8\pi}{\alpha}\frac{\Delta m_{\pi}}{m_\pi}\approx 113$.} 
\begin{equation}\label{eq:pionsumrulephyspheno}
\frac{8\pi}{\alpha}\frac{\Delta m_{\pi}}{m_\pi}
=-I\;\frac{3\left(  1  +\delta_{\chi}^{(2)}\right) }{F_{\pi}^2m_{\pi}^2}
 +\delta_{\chi}^{(1)} \, ,
\end{equation}
where
\begin{equation}\begin{aligned}\label{eq:Iint}
I&=\frac{1}{\pi}\int^{s_0}_{s_{th}}dk^2\mathrm{Im}\Pi(k^2) \, k^2\ln\frac{k^2}{s_0}+ \delta_{\mathrm{DV}}(s_0)+F_{\pi}^2 m_\pi^2\, \ln\frac{s_0}{\mu_{\mathrm{SD}}^2}\, ,\\
\delta_{\mathrm{DV}}(s_0)&=\frac{1}{\pi}\int^{\infty}_{s_{0}}dk^2\mathrm{Im}\Pi(k^2) \, k^2\ln\frac{k^2}{s_0} \, ,
\end{aligned}
\end{equation}
and, at $\xi = 1, \,\,  \mu_{\textrm{SD}}= 1 \, \textrm{GeV}, \, \, \nu_{\chi}= 0.77 \, \textrm{GeV}$, which will be our matching point from now on,
\begin{equation}
\begin{aligned}
\delta_{\chi}^{(1)}&=  3  
   \ln \frac{\nu_\chi^2}{m_{\pi}^2}+\xi  
   \ln \frac{\mu^2_{\mathrm{SD}}}{\nu _\chi ^2}+5+\xi -32\pi^2  \left(
  -4 
    K_{11}^r
   +2 K_{3}^r
   - K_{4}^r
   \right) =17.1 \pm 2.2\,  
    \, , \\
\delta_{\chi}^{(2)} &=  \frac{m_\pi^2}{(4\pi F_{\pi})^2}\left(\ln \frac{\nu _\chi ^2}{m_\pi^2}-1
\right)=0.035 \, ,
\end{aligned}
\end{equation}
dominated by the large infrared pion mass logarithms. 

\section{The pion sum rule combined with hadronic tau-decay data}
\label{sec:pionsrpheno}

\begin{figure}
    \centering
    \includegraphics[width=0.9\linewidth]{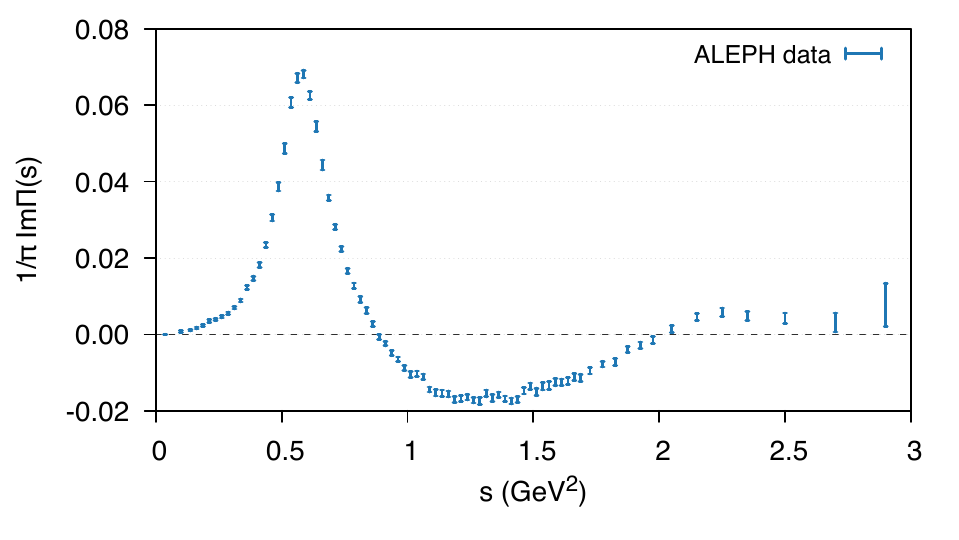}
    \caption{$\mathrm{Im}\Pi$ as obtained from ALEPH hadronic tau decay data.}
    \label{fig:vminusa}
\end{figure}

The correlator $\Pi$ entering into $I$ in Eq.~(\ref{eq:pionsumrulephyspheno}) is, in the isospin limit,\footnote{Potential differences emerging from using this approximation are second order in isospin breaking, and, thus, consistently neglected. While at the working order it is indifferent (except for the difference), considering that we are working with charged currents including the charged pion pole, we will be using as input for the pion decay constants and masses the charged ones.} half the charged correlator $\Pi_{ud}$, whose imaginary part can be extracted from hadronic $\tau$ data up to $k^2=m_{\tau}^2$, see e.g.~Ref.~\cite{Braaten:1991qm}, or Ref.~\cite{Rodriguez-Sanchez:2025nsm} for a more recent pedagogical introduction. We will be using the publicly available ALEPH spectral functions of Ref.~\cite{Davier:2013sfa}, and perform the calculation of the dispersive integral following Ref.~\cite{Pich:2016bdg}. The corresponding experimental spectral function is displayed in Fig.~\ref{fig:vminusa}.

Therefore, the most natural phenomenological application of our main result, i.e. Eq.~(\ref{eq:pionsumrulephyspheno}), is
using the experimental difference of pion masses to predict $I$, $I_{\Delta m_{\pi}^{\mathrm{exp}}}$, which can then be compared to what one obtains using experimental data. One finds, including corrections linear in the quark masses (NLO),
\begin{equation}\label{eq:NLOpred}
I^{\mathrm{NLO}}_{\Delta m_{\pi}^{\mathrm{exp}}}= -0.00514(12) \, \mathrm{GeV}^4 .
\end{equation}
Notice the significant shift with respect the naive chiral limit prediction obtained by taking $\delta^{(1)}_{\chi}=\delta^{(2)}_{\chi}=0$, $I^{\mathrm{LO}}_{\Delta m_{\pi}^{\mathrm{exp}}}\approx -0.0063\, \mathrm{GeV}^4$.

\begin{figure}[tb]
    \centering
    \includegraphics[width=0.9\linewidth]{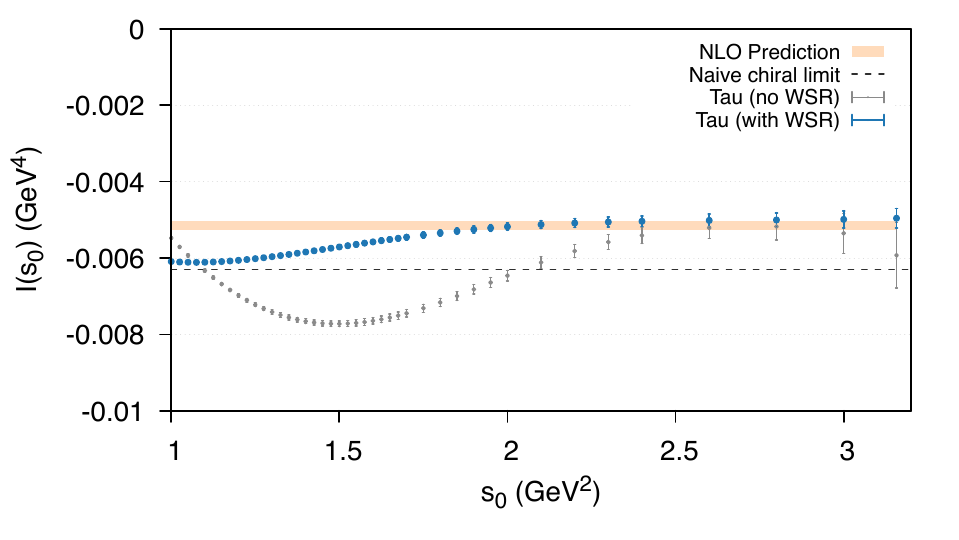}
    \caption{$I(s_0)$ neglecting duality violations, using or not the Weinberg Sum Rules, compared to the predicted NLO value obtained in this work.}
    \label{fig:Is0}
\end{figure}

For the tau-based evaluation of $I$, one has to use
\begin{equation}\begin{aligned}
I&=\frac{1}{\pi}\int^{s_0}_{s_{th}}dk^2\mathrm{Im}\Pi(k^2) \, k^2\ln\frac{k^2}{s_0}+ \delta_{\mathrm{DV}}(s_0)+F_{\pi}^2 m_\pi^2\, \ln\frac{s_0}{\mu_{\mathrm{SD}}^2}\, ,
\end{aligned}
\end{equation}
and a challenge is the fact that one does not really know how to compute the $\delta_{\mathrm{DV}}(s_0)$ component from first principles. They are however known to go to zero very fast at large $s_0$, typically exponentially, so the contributions are dominated by the region close to $s_0$. In order to find a tau-based $I$ where the $\mathrm{DV}$ is not large, we may combine it with the Weinberg Sum Rules, namely, \cite{Weinberg:1967kj}
\begin{equation}
\delta I(\delta_1,\delta_2)=\frac{1}{\pi}\int^{s_0}_{s_{th}}dk^2\left[\mathrm{Im}\Pi(k^2) (\delta_1 s_0+\delta_2 k^2 )\right] -F_{\pi}^2(\delta_1 s_0+\delta_{2}M_{\pi}^2)+\delta'_{\mathrm{DV}}(\delta_1,\delta_2,s_0)=0 \, ,
\end{equation}
which can be easily derived by integrating, with the corresponding weights, along the circuit $\mathcal{C}_1$ in Fig.~\ref{fig:circuit}. Here $\delta'_{\mathrm{DV}}(\delta_1,\delta_2,s_0)$ denotes the corresponding integral from $s_0$ up to infinity, analogous to $\delta_{\mathrm{DV}}$ in Eq.~(\ref{eq:Iint}).
 We may thus take
\begin{equation}
\begin{aligned}
I=I(\delta_1,\delta_2,s_0)&=\frac{1}{\pi}\int^{s_0}_{s_{th}}dk^2\left[\mathrm{Im}\Pi(k^2) \left(\delta_1 s_0+\delta_2 k^2 +k^2\ln\frac{k^2}{s_0}\right)\right] \\
&-F_{\pi}^2\left(\delta_1 s_0+M_{\pi}^2\left(\delta_2-\ln\frac{s_0}{\mu_{\textrm{SD}}^2}\right)\right)+\delta_{\mathrm{DV}}(\delta_1,\delta_2,s_0)=0 \, ,
\end{aligned}
\end{equation}
where
\begin{equation}
\delta_{\mathrm{DV}}(\delta_1,\delta_2,s_0)=\frac{1}{\pi}\int^{\infty}_{s_0} dk^2\left[\mathrm{Im}\Pi_{\mathrm{DV}}(k^2) \left(\delta_1 s_0+\delta_2 k^2 +k^2\ln\frac{k^2}{s_0}\right)\right] \, .
\end{equation}
Up to unwanted DVs, any $\delta_1$ and $\delta_2$ choices return $I$. Two optimal choices to further suppress DVs and also experimental uncertainties are $\delta_1=\delta_2=0$ (vanishing weight at $s_0$, actually not using the Weinberg Sum Rules) and, even better (vanishing weight and its derivative at $s_0$), $\delta_1=-\delta_2=1$ (therefore using the Weinberg sum rules). Taking as reference value $s_0=2.8\,\mathrm{GeV}^2$, large enough so that we can neglect duality violations in the integrals but below the point where experimental uncertainties becomes very large (see Fig.~\ref{fig:vminusa}), we find
\begin{equation}
\begin{aligned}
I^{\tau}(\mathrm{no\; WSR})&=-0.00517(35) \quad \, ,\\
I^{\tau}(\mathrm{with\; WSR})&=-0.00500(19) \quad \, .
\end{aligned}
\end{equation}
This is in excellent agreement with Eq.~(\ref{eq:NLOpred}). In order to illustrate how $s_0=2.8\,\mathrm{GeV}^2$ is large enough to have converged to the quark-hadron duality regime, especially for the second weight, we show in Fig.~\ref{fig:Is0} the corresponding integrals up to $s_0$, where one can see how they have converged both to the expected $s_0$-independent value and to each other well below the tau mass.\footnote{One can also check using the different models for duality violations in the literature, e.g. see Refs.~\cite{Boito:2014sta,Gonzalez-Alonso:2016ndl,Pich:2022tca}, that these residual contributions are expected to be below the quoted uncertainties.}

Equivalently, one may instead take $I$ from tau data to predict the relative pion mass difference, whose experimental value is $\frac{\Delta m_{\pi}}{m_{\pi}}=0.03291\dots$. The corresponding values at $s_0=2.8\, \mathrm{GeV}^2$ are
\begin{equation}
\begin{aligned}
\left[\frac{\Delta m_{\pi}}{m_{\pi}}\right]^{\tau}(\mathrm{no \; WSR})&=0.0331(20) \, , \, \\
\left[\frac{\Delta m_{\pi}}{m_{\pi}}\right]^{\tau}(\mathrm{with \; WSR})&=0.0322(12) \, , \,
\end{aligned}
\end{equation}
and at any other $s_0$, together with the experimental value. Notice again that for the NLO piece $\delta _\chi^{(1)}$ not enhanced by the pion mass logarithm, we have used the estimate of Eq.~(\ref{eq:Kestimates}).
\begin{figure}[tb]
    \centering
    \includegraphics[width=0.9\linewidth]{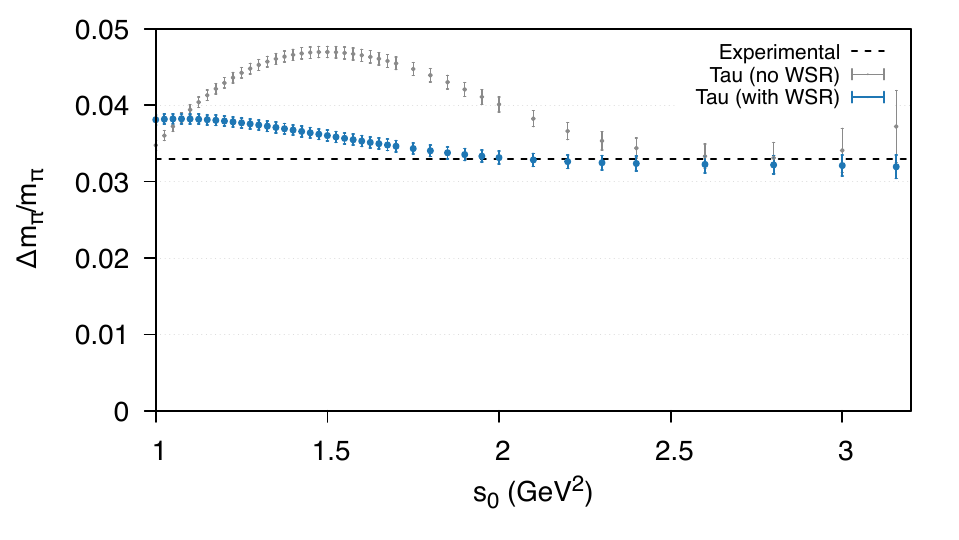}
    \caption{The ratio $\frac{\Delta m_{\pi}}{m_{\pi}}$ from $I_\tau(s_0)$ neglecting duality violations, using or not the Weinberg Sum Rules, compared to the experimental value.}
    \label{fig:pionmassdifference}
\end{figure}

Incorporating the NLO corrections in the chiral counting to the pion sum rule has thus improved the agreement with experimental data. Let us however finish this section by warning that there may be higher-order corrections in the chiral counting above the per-cent level and thus sizable, for example corresponding to disconnected diagrams in QCD  involving quadratic strange-quark mass corrections. 

\section{Conclusions}
In this paper we have generalized a well-known sum rule from Refs.~\cite{Das:1967it,Moussallam:1997xx} relating the pion mass difference between charged and neutral particles to experimental data. Combining different nonperturbative techniques, we have generalized the pion sum rule as a finite-energy sum rule incorporating also quark mass mass corrections. 
The chiral perturbation theory estimates used here are valid through next-to-leading order, or one loop, in the chiral counting in the presence of electromagnetism (order $p^4$, $e^2 p^4$)~\cite{Urech:1994hd}. Extending beyond this order is non-trivial, as it first of all would require deriving the additional Lagrangian of order $e^2 p^4$ beyond Ref.~\cite{Urech:1994hd} and estimating new low-energy constants appearing. Once known, the Lagrangian would have to be used. An additional complication is that more of the $K_i$ could appear, from vertex insertions in one-loop diagrams. Nevertheless, 
we have shown that the current corrections are sizable and how incorporating them improves the precision of this dispersion relation, which relates the difference of masses of the two lightest hadrons, mainly of electromagnetic origin, to hadronic tau-decay data~\cite{Davier:2013sfa}, an electroweak process. Our generalized sum rule can in the future be used for dispersive studies beyond the chiral limit, such as including isospin-breaking effects generally needed for high-precision predictions. Moreover, it can provide an important ingredient for future improved determinations of some of the poorly known electromagnetic low-energy constants of chiral perturbation theory. 

\section*{Acknowledgements}
This work has been supported by 
the Spanish Government (Agencia Estatal de Investigaci\'on MCIN/AEI/10.13039/501100011033) Grant No. PID2023-146220NB-I00 and by the Generalitat Valenciana (Spain) through the plan GenT program (CIDEIG/2023/12). N.~H.-T.~is supported by the UK Research and Innovation, Science and Technology Facilities Council, grant number UKRI2426. We thank Antonio Pich for valuable input and carefully reading the manuscript. N.~H.-T.~and A.~R.-S.~thank Lund University for hosting a research visit during which a significant part of the work was undertaken. 

\appendix

\bibliographystyle{JHEP}
\bibliography{refs}

@article{Gasser:1983yg,
    author = "Gasser, J. and Leutwyler, H.",
    title = "{Chiral Perturbation Theory to One Loop}",
    reportNumber = "CERN-TH-3689",
    doi = "10.1016/0003-4916(84)90242-2",
    journal = "Annals Phys.",
    volume = "158",
    pages = "142",
    year = "1984"
}

@article{Gasser:1984gg,
    author = "Gasser, J. and Leutwyler, H.",
    title = "{Chiral Perturbation Theory: Expansions in the Mass of the Strange Quark}",
    reportNumber = "CERN-TH-3798",
    doi = "10.1016/0550-3213(85)90492-4",
    journal = "Nucl. Phys. B",
    volume = "250",
    pages = "465--516",
    year = "1985"
}

@article{Witten:1983tw,
    author = "Witten, Edward",
    title = "{Global Aspects of Current Algebra}",
    reportNumber = "PRINT-83-0262 (PRINCETON)",
    doi = "10.1016/0550-3213(83)90063-9",
    journal = "Nucl. Phys. B",
    volume = "223",
    pages = "422--432",
    year = "1983"
}

@article{WESS197195,
title = {Consequences of anomalous ward identities},
journal = {Physics Letters B},
volume = {37},
number = {1},
pages = {95-97},
year = {1971},
issn = {0370-2693},
doi = {https://doi.org/10.1016/0370-2693(71)90582-X},
url = {https://www.sciencedirect.com/science/article/pii/037026937190582X},
author = {J. Wess and B. Zumino}
}

@article{Ebertshauser:2001nj,
    author = "Ebertshauser, T. and Fearing, H. W. and Scherer, S.",
    title = "{The Anomalous chiral perturbation theory meson Lagrangian to order $p^6$ revisited}",
    eprint = "hep-ph/0110261",
    archivePrefix = "arXiv",
    reportNumber = "MKPH-T-01-22, TRI-PP-01-34",
    doi = "10.1103/PhysRevD.65.054033",
    journal = "Phys. Rev. D",
    volume = "65",
    pages = "054033",
    year = "2002"
}

@article{Urech:1994hd,
    author = "Urech, Res",
    title = "{Virtual photons in chiral perturbation theory}",
    eprint = "hep-ph/9405341",
    archivePrefix = "arXiv",
    reportNumber = "BUTP-94-9",
    doi = "10.1016/0550-3213(95)90707-N",
    journal = "Nucl. Phys. B",
    volume = "433",
    pages = "234--254",
    year = "1995"
}

@article{Neufeld:1995mu,
    author = "Neufeld, H. and Rupertsberger, H.",
    title = "{The Electromagnetic interaction in chiral perturbation theory}",
    eprint = "hep-ph/9506448",
    archivePrefix = "arXiv",
    reportNumber = "UWTHPH-1995-18",
    doi = "10.1007/s002880050156",
    journal = "Z. Phys. C",
    volume = "71",
    pages = "131--138",
    year = "1996"
}

@article{Pich:2018ltt,
    author = "Pich, Antonio",
    editor = "Davidson, Sacha and Gambino, Paolo and Laine, Mikko and Neubert, Matthias and Salomon, Christophe",
    title = "{Effective Field Theory with Nambu-Goldstone Modes}",
    eprint = "1804.05664",
    archivePrefix = "arXiv",
    primaryClass = "hep-ph",
    reportNumber = "IFIC/18-13, FTUV/18-0415, IFIC-18-13, FTUV-18-0415",
    doi = "10.1093/oso/9780198855743.003.0003",
    month = "4",
    year = "2018"
}

@article{ECKER1989311,
title = {The role of resonances in chiral perturbation theory},
journal = {Nuclear Physics B},
volume = {321},
number = {2},
pages = {311-342},
year = {1989},
issn = {0550-3213},
doi = {https://doi.org/10.1016/0550-3213(89)90346-5},
url = {https://www.sciencedirect.com/science/article/pii/0550321389903465},
author = {G. Ecker and J. Gasser and A. Pich and E. {De Rafael}}
}

@article{Bijnens:2001bb,
    author = "Bijnens, J. and Girlanda, L. and Talavera, P.",
    title = "{The anomalous chiral Lagrangian of order $p^6$}",
    eprint = "hep-ph/0110400",
    archivePrefix = "arXiv",
    reportNumber = "LU-TP-01-34, DFPD-01-TH-25, CPT-2001-P4256",
    doi = "10.1007/s100520100887",
    journal = "Eur. Phys. J. C",
    volume = "23",
    pages = "539--544",
    year = "2002"
}

@article{PhysRevD.53.315,
  title = {Extension of the chiral perturbation theory meson Lagrangian to order ${\mathit{p}}^{6}$},
  author = {Fearing, H. W. and Scherer, S.},
  journal = {Phys. Rev. D},
  volume = {53},
  issue = {1},
  pages = {315--348},
  numpages = {0},
  year = {1996},
  month = {Jan},
  publisher = {American Physical Society},
  doi = {10.1103/PhysRevD.53.315},
  url = {https://link.aps.org/doi/10.1103/PhysRevD.53.315}
}

@article{Bijnens:1999sh,
      author         = "Bijnens, Johan and Colangelo, Gilberto and Ecker,
                        Gerhard",
      title          = "{The Mesonic chiral Lagrangian of order $p^6$}",
      journal        = "JHEP",
      volume         = "02",
      year           = "1999",
      pages          = "020",
      doi            = "10.1088/1126-6708/1999/02/020",
      eprint         = "hep-ph/9902437",
      archivePrefix  = "arXiv",
      primaryClass   = "hep-ph",
      reportNumber   = "LU-TP-99-02, UWTHPH-1999-02, ZU-TH-9-99",
      SLACcitation   = "%%CITATION = HEP-PH/9902437;%%"
}

@article{Bijnens:2006mk,
    author = "Bijnens, Johan and Danielsson, Niclas",
    title = "{Electromagnetic Corrections in Partially Quenched Chiral Perturbation Theory}",
    eprint = "hep-lat/0610127",
    archivePrefix = "arXiv",
    reportNumber = "LU-TP-06-38",
    doi = "10.1103/PhysRevD.75.014505",
    journal = "Phys. Rev. D",
    volume = "75",
    pages = "014505",
    year = "2007"
}

@mastersthesis{Weber:2008,
author = "Weber, Johannes",
title = "{Mesonic Lagrangians and Anomalous Processes}",
school = "{Mainz Institut f\"ur Kernphysik}",
url = "{https://wwwth.kph.uni-mainz.de/files/2010/03/Dipl\_J\_Weber.pdf}",
year= "2008"
 }

@article{Bijnens:2018lez,
    author = "Bijnens, Johan and Hermansson-Truedsson, Nils and Wang, Si",
    title = "{The order p$^{8}$ mesonic chiral Lagrangian}",
    eprint = "1810.06834",
    archivePrefix = "arXiv",
    primaryClass = "hep-ph",
    reportNumber = "LU TP 18-34",
    doi = "10.1007/JHEP01(2019)102",
    journal = "JHEP",
    volume = "01",
    pages = "102",
    year = "2019"
}

@article{Hermansson-Truedsson:2020rtj,
    author = "Hermansson-Truedsson, Nils",
    title = "{Chiral Perturbation Theory at NNNLO}",
    eprint = "2006.01430",
    archivePrefix = "arXiv",
    primaryClass = "hep-ph",
    doi = "10.3390/sym12081262",
    journal = "Symmetry",
    volume = "12",
    number = "8",
    pages = "1262",
    year = "2020"
}

@article{Bijnens:2023hyv,
    author = "Bijnens, Johan and Hermansson-Truedsson, Nils and Ruiz-Vidal, Joan",
    title = "{The anomalous chiral Lagrangian at order p$^{8}$}",
    eprint = "2310.20547",
    archivePrefix = "arXiv",
    primaryClass = "hep-ph",
    doi = "10.1007/JHEP01(2024)009",
    journal = "JHEP",
    volume = "01",
    pages = "009",
    year = "2024"
}

@article{Bijnens:2014lea,
    author = "Bijnens, Johan and Ecker, Gerhard",
    title = "{Mesonic low-energy constants}",
    eprint = "1405.6488",
    archivePrefix = "arXiv",
    primaryClass = "hep-ph",
    reportNumber = "LU-TP-14-16, UWTHPH-2014-10",
    doi = "10.1146/annurev-nucl-102313-025528",
    journal = "Ann. Rev. Nucl. Part. Sci.",
    volume = "64",
    pages = "149--174",
    year = "2014"
}

@article{Gasser:2003hk,
    author = "Gasser, J. and Rusetsky, A. and Scimemi, I.",
    title = "{Electromagnetic corrections in hadronic processes}",
    eprint = "hep-ph/0305260",
    archivePrefix = "arXiv",
    reportNumber = "BUTP-03-06, ECT-03-05",
    doi = "10.1140/epjc/s2003-01383-1",
    journal = "Eur. Phys. J. C",
    volume = "32",
    pages = "97--114",
    year = "2003"
}

@article{Agadjanov:2013lra,
    author = "Agadjanov, Andria and Agadjanov, Dimitri and Khelashvili, Anzor and Rusetsky, Akaki",
    title = "{Generating functional for mesonic ChPT with virtual photons in a general covariant gauge}",
    eprint = "1307.1451",
    archivePrefix = "arXiv",
    primaryClass = "hep-ph",
    doi = "10.1140/epja/i2013-13120-x",
    journal = "Eur. Phys. J. A",
    volume = "49",
    pages = "120",
    year = "2013"
}

@article{Bijnens:1996kk,
    author = "Bijnens, Johan and Prades, Joaquim",
    title = "{Electromagnetic corrections for pions and kaons: Masses and polarizabilities}",
    eprint = "hep-ph/9610360",
    archivePrefix = "arXiv",
    reportNumber = "FTUV-96-69, IFIC-96-78, NORDITA-96-70-N-P",
    doi = "10.1016/S0550-3213(97)00107-7",
    journal = "Nucl. Phys. B",
    volume = "490",
    pages = "239--271",
    year = "1997"
}

@article{Bijnens:1993ae,
    author = "Bijnens, Johan",
    title = "{Violations of Dashen's theorem}",
    eprint = "hep-ph/9302217",
    archivePrefix = "arXiv",
    reportNumber = "NORDITA-93-15-N-P",
    doi = "10.1016/0370-2693(93)90089-Z",
    journal = "Phys. Lett. B",
    volume = "306",
    pages = "343--349",
    year = "1993"
}

@article{Moussallam:1997xx,
    author = "Moussallam, B.",
    title = "{A Sum rule approach to the violation of Dashen's theorem}",
    eprint = "hep-ph/9701400",
    archivePrefix = "arXiv",
    reportNumber = "IPNO-TH-97-03",
    doi = "10.1016/S0550-3213(97)00464-1",
    journal = "Nucl. Phys. B",
    volume = "504",
    pages = "381--414",
    year = "1997"
}

@article{Braaten:1991qm,
    author = "Braaten, E. and Narison, Stephan and Pich, A.",
    title = "{QCD analysis of the tau hadronic width}",
    reportNumber = "CERN-TH-6070-91, NUHEP-TH-91-8, PM-91-8",
    doi = "10.1016/0550-3213(92)90267-F",
    journal = "Nucl. Phys. B",
    volume = "373",
    pages = "581--612",
    year = "1992"
}

@article{Amoros:1999dp,
    author = "Amoros, Gabriel and Bijnens, Johan and Talavera, P.",
    title = "{Two point functions at two loops in three flavor chiral perturbation theory}",
    eprint = "hep-ph/9907264",
    archivePrefix = "arXiv",
    reportNumber = "LU-TP-99-15",
    doi = "10.1016/S0550-3213(99)00674-4",
    journal = "Nucl. Phys. B",
    volume = "568",
    pages = "319--363",
    year = "2000"
}

@article{Bijnens:1994ci,
    author = "Bijnens, Johan and Prades, Joaquim and de Rafael, Eduardo",
    title = "{Light quark masses in QCD}",
    eprint = "hep-ph/9411285",
    archivePrefix = "arXiv",
    reportNumber = "CPT-94-PE-3097, NORDITA-94-62",
    doi = "10.1016/0370-2693(95)00105-T",
    journal = "Phys. Lett. B",
    volume = "348",
    pages = "226--238",
    year = "1995"
}

@article{Das:1967it,
    author = "Das, T. and Guralnik, G. S. and Mathur, V. S. and Low, F. E. and Young, J. E.",
    title = "{Electromagnetic mass difference of pions}",
    doi = "10.1103/PhysRevLett.18.759",
    journal = "Phys. Rev. Lett.",
    volume = "18",
    pages = "759--761",
    year = "1967"
}

@article{Ecker:1988te,
    author = "Ecker, G. and Gasser, J. and Pich, A. and de Rafael, E.",
    title = "{The Role of Resonances in Chiral Perturbation Theory}",
    reportNumber = "CERN-TH-5185/88, UWThPh-1988-29, BUTP-88/18, CPT-88/PE-2158",
    doi = "10.1016/0550-3213(89)90346-5",
    journal = "Nucl. Phys. B",
    volume = "321",
    pages = "311--342",
    year = "1989"
}

@article{Shifman:1978bx,
    author = "Shifman, Mikhail A. and Vainshtein, A. I. and Zakharov, Valentin I.",
    title = "{QCD and Resonance Physics. Theoretical Foundations}",
    reportNumber = "ITEP-73-1978, ITEP-80-1978",
    doi = "10.1016/0550-3213(79)90022-1",
    journal = "Nucl. Phys. B",
    volume = "147",
    pages = "385--447",
    year = "1979"
}

@article{Chibisov:1996wf,
    author = "Chibisov, Boris and Dikeman, R. David and Shifman, Mikhail A. and Uraltsev, N.",
    title = "{Operator product expansion, heavy quarks, QCD duality and its violations}",
    eprint = "hep-ph/9605465",
    archivePrefix = "arXiv",
    reportNumber = "TPI-MINN-96-05-T, UMN-TH-1426-96, CERN-TH-96-113",
    doi = "10.1142/S0217751X97001316",
    journal = "Int. J. Mod. Phys. A",
    volume = "12",
    pages = "2075--2133",
    year = "1997"
}

@article{Cata:2008ye,
    author = "Cata, Oscar and Golterman, Maarten and Peris, Santi",
    title = "{Unraveling duality violations in hadronic tau decays}",
    eprint = "0803.0246",
    archivePrefix = "arXiv",
    primaryClass = "hep-ph",
    doi = "10.1103/PhysRevD.77.093006",
    journal = "Phys. Rev. D",
    volume = "77",
    pages = "093006",
    year = "2008"
}

@article{Gonzalez-Alonso:2010kpl,
    author = "Gonzalez-Alonso, Martin and Pich, Antonio and Prades, Joaquim",
    title = "{Violation of Quark-Hadron Duality and Spectral Chiral Moments in QCD}",
    eprint = "1001.2269",
    archivePrefix = "arXiv",
    primaryClass = "hep-ph",
    reportNumber = "CAFPE-122-09, FTUV-10-0113, IFIC-09-68, UG-FT-252-09",
    doi = "10.1103/PhysRevD.81.074007",
    journal = "Phys. Rev. D",
    volume = "81",
    pages = "074007",
    year = "2010"
}

@article{Boito:2017cnp,
    author = "Boito, Diogo and Caprini, Irinel and Golterman, Maarten and Maltman, Kim and Peris, Santiago",
    title = "{Hyperasymptotics and quark-hadron duality violations in QCD}",
    eprint = "1711.10316",
    archivePrefix = "arXiv",
    primaryClass = "hep-ph",
    doi = "10.1103/PhysRevD.97.054007",
    journal = "Phys. Rev. D",
    volume = "97",
    number = "5",
    pages = "054007",
    year = "2018"
}

@article{Cirigliano:2019cpi,
    author = "Cirigliano, V. and Gisbert, H. and Pich, A. and Rodr{\'\i}guez-S{\'a}nchez, A.",
    title = "{Isospin-violating contributions to  $\epsilon'/\epsilon$}",
    eprint = "1911.01359",
    archivePrefix = "arXiv",
    primaryClass = "hep-ph",
    reportNumber = "LU TP/19-51, IFIC/19-46, DO-TH/19-23",
    doi = "10.1007/JHEP02(2020)032",
    journal = "JHEP",
    volume = "02",
    pages = "032",
    year = "2020"
}

@article{Davier:2013sfa,
    author = {Davier, Michel and H{\"o}cker, Andreas and Malaescu, Bogdan and Yuan, Chang-Zheng and Zhang, Zhiqing},
    title = "{Update of the ALEPH non-strange spectral functions from hadronic $\tau$ decays}",
    eprint = "1312.1501",
    archivePrefix = "arXiv",
    primaryClass = "hep-ex",
    reportNumber = "LAL-13-390",
    doi = "10.1140/epjc/s10052-014-2803-9",
    journal = "Eur. Phys. J. C",
    volume = "74",
    number = "3",
    pages = "2803",
    year = "2014"
}

@article{Weinberg:1967kj,
    author = "Weinberg, Steven",
    title = "{Precise relations between the spectra of vector and axial vector mesons}",
    doi = "10.1103/PhysRevLett.18.507",
    journal = "Phys. Rev. Lett.",
    volume = "18",
    pages = "507--509",
    year = "1967"
}

@article{Pich:2022tca,
    author = "Pich, Antonio and Rodr{\'\i}guez-S{\'a}nchez, Antonio",
    title = "{Violations of quark-hadron duality in low-energy determinations of {\ensuremath{\alpha}}$_{s}$}",
    eprint = "2205.07587",
    archivePrefix = "arXiv",
    primaryClass = "hep-ph",
    doi = "10.1007/JHEP07(2022)145",
    journal = "JHEP",
    volume = "07",
    pages = "145",
    year = "2022"
}

@article{Boito:2014sta,
    author = "Boito, Diogo and Golterman, Maarten and Maltman, Kim and Osborne, James and Peris, Santiago",
    title = "{Strong coupling from the revised ALEPH data for hadronic $\tau$ decays}",
    eprint = "1410.3528",
    archivePrefix = "arXiv",
    primaryClass = "hep-ph",
    doi = "10.1103/PhysRevD.91.034003",
    journal = "Phys. Rev. D",
    volume = "91",
    number = "3",
    pages = "034003",
    year = "2015"
}

@article{Gonzalez-Alonso:2016ndl,
    author = "Gonz{\'a}lez-Alonso, Martin and Pich, Antonio and Rodr{\'\i}guez-S{\'a}nchez, Antonio",
    title = "{Updated determination of chiral couplings and vacuum condensates from hadronic $\tau$ decay data}",
    eprint = "1602.06112",
    archivePrefix = "arXiv",
    primaryClass = "hep-ph",
    doi = "10.1103/PhysRevD.94.014017",
    journal = "Phys. Rev. D",
    volume = "94",
    number = "1",
    pages = "014017",
    year = "2016"
}

@article{FlavourLatticeAveragingGroupFLAG:2024oxs,
    author = "Aoki, Y. and others",
    collaboration = "Flavour Lattice Averaging Group (FLAG)",
    title = "{FLAG Review 2024}",
    eprint = "2411.04268",
    archivePrefix = "arXiv",
    primaryClass = "hep-lat",
    reportNumber = "CERN-TH-2024-192, FERMILAB-PUB-24-0785-T",
    month = "11",
    year = "2024"
}

@article{Ananthanarayan:2000ht,
    author = "Ananthanarayan, B. and Colangelo, G. and Gasser, J. and Leutwyler, H.",
    title = "{Roy equation analysis of pi pi scattering}",
    eprint = "hep-ph/0005297",
    archivePrefix = "arXiv",
    reportNumber = "IISC-CTS-12-99, ZU-TH-10-00, BUTP-99-33",
    doi = "10.1016/S0370-1573(01)00009-6",
    journal = "Phys. Rept.",
    volume = "353",
    pages = "207--279",
    year = "2001"
}

@article{Aliberti:2025beg,
    author = "Aliberti, R. and others",
    title = "{The anomalous magnetic moment of the muon in the Standard Model: an update}",
    eprint = "2505.21476",
    archivePrefix = "arXiv",
    primaryClass = "hep-ph",
    reportNumber = "CERN-TH-2025-101, FERMILAB-PUB-25-0344-T, INT-PUB-25-015, IPARCOS-UCM-25-029, KEK Preprint 2025-22, LTH 1403, MITP-25-037, UWThPh 2025-15, UWThPh
  2025-15, ZU-TH 37/25, IPARCOS-UCM-25-029",
    doi = "10.1016/j.physrep.2025.08.002",
    journal = "Phys. Rept.",
    volume = "1143",
    pages = "1--158",
    year = "2025"
}

@article{Aoyama:2020ynm,
    author = "Aoyama, T. and others",
    title = "{The anomalous magnetic moment of the muon in the Standard Model}",
    eprint = "2006.04822",
    archivePrefix = "arXiv",
    primaryClass = "hep-ph",
    reportNumber = "FERMILAB-PUB-20-207-T, INT-PUB-20-021, KEK Preprint 2020-5,
  MITP/20-028, KEK Preprint 2020-5, MITP/20-028, CERN-TH-2020-075, IFT-UAM/CSIC-20-74, LMU-ASC 18/20, LTH 1234,
  LU TP 20-20, LTH 1234, LU TP 20-20, MAN/HEP/2020/003, PSI-PR-20-06, UWThPh 2020-14, ZU-TH 18/20",
    doi = "10.1016/j.physrep.2020.07.006",
    journal = "Phys. Rept.",
    volume = "887",
    pages = "1--166",
    year = "2020"
}

@article{Pich:2016bdg,
    author = "Pich, Antonio and Rodr{\'\i}guez-S{\'a}nchez, Antonio",
    title = "{Determination of the QCD coupling from ALEPH $\tau$ decay data}",
    eprint = "1605.06830",
    archivePrefix = "arXiv",
    primaryClass = "hep-ph",
    reportNumber = "IFIC-16-13, FTUV-16-0522",
    doi = "10.1103/PhysRevD.94.034027",
    journal = "Phys. Rev. D",
    volume = "94",
    number = "3",
    pages = "034027",
    year = "2016"
}

@incollection{Rodriguez-Sanchez:2025nsm,
    author = "Rodr{\'\i}guez-S{\'a}nchez, Antonio",
    title = "{Hadronic tau decays}",
booktitle = {To appear in the Encyclopedia of Particle Physics},
publisher = {Elsevier},
isbn = {978-0-12-803581-8},
doi = {https://doi.org/10.1016/B978-0-443-26598-3.00027-4},
url = {https://www.sciencedirect.com/science/article/pii/B9780443265983000274},
    eprint = "2504.21732",
    archivePrefix = "arXiv",
    primaryClass = "hep-ph",
year = {2025}
}

\end{document}